\documentclass[english,latin9]{aa}

\usepackage[T1]{fontenc}
\usepackage{color}
\usepackage{amsmath}
\usepackage{graphicx}
\usepackage{babel}
\usepackage{multirow}
\usepackage{siunitx}
\usepackage{multirow}
\usepackage{graphicx}
\usepackage{txfonts}
\usepackage{natbib,twoopt}
\usepackage[breaklinks=true]{hyperref}
\bibpunct{(}{)}{;}{a}{}{,}

\begin{document} 

\title{The Global sphere reconstruction (GSR)}
\subtitle{Demonstrating an independent implementation\\ of the astrometric core solution for Gaia}

\author{A. Vecchiato\inst{1} \and 
   		B. Bucciarelli\inst{1} \and 
		M.~G. Lattanzi\inst{1} \and
        U. Becciani\inst{2} \and
        L. Bianchi\inst{3} \and
        U. Abbas\inst{1} \and
        E. Sciacca\inst{2} \and
		R. Messineo\inst{4} \and 
		R. De March\inst{4}}

\institute{INAF - Astrophysical Observatory of Torino, Pino Torinese\\
		   \email{alberto.vecchiato@inaf.it}
	\and   INAF - Astrophysical Observatory of Catania
	\and   EURIX, Torino
	\and   ALTEC, Torino
}

\offprints{alberto.vecchiato@inaf.it}

\date{Received ...; Accepted...}

\abstract{The Gaia ESA mission will estimate the astrometric and physical data of more than one billion objects, providing the largest and most precise catalog of absolute astrometry in the history of Astronomy. The core of this process, the so-called global sphere reconstruction, is represented by the reduction of a subset of these objects which will be used to define the celestial reference frame. As the Hipparcos mission showed, and as is inherent to all kinds of absolute measurements, possible errors in the data reduction can hardly be identified from the catalog, thus potentially introducing systematic errors in all derived work.}
{Following up on the lessons learned from Hipparcos, our aim is thus to develop an independent sphere reconstruction method that contributes to guarantee the quality of the astrometric results without fully reproducing the main processing chain.}
{Indeed, given the unfeasibility of a complete replica of the data reduction pipeline, an astrometric verification unit (AVU) was instituted by the Gaia Data Processing and Analysis Consortium (DPAC). One of its jobs is to implement and operate an independent global sphere reconstruction (GSR), parallel to the baseline one (AGIS, namely Astrometric Global Iterative Solution) but limited to the primary stars and for validation purposes, to compare the two results, and to report on any significant differences.}
{Tests performed on simulated data show that GSR is able to reproduce at the sub-$\mu$as level the results of the AGIS demonstration run presented in Lindegren et al. (2012).}
{Further development is ongoing to improve on the treatment of real data and on the software modules that compare the AGIS and GSR solutions to identify possible discrepancies above the tolerance level set by the accuracy of the Gaia catalog.}


\maketitle

\section{Introduction\label{sec:Introduction}}
The main goal of the Gaia mission, an ESA (European Space Agency) satellite launched in December 2013, is the production of a five-parameter astrometric catalog (i.e., including positions, parallaxes and the two components of the proper motions) at the 10 to 1000~$\mu$arcsecond-level ($\mu$as) for about one billion stars of our Galaxy in the magnitude range from 3 to 20.7 \citep{2016A&A...595A...1G}.

To this end, the satellite has been designed as a scanning telescope that sweeps continuously and repeatedly the entire celestial sphere during the five years of its foreseen mission duration. The target accuracy can be reached by averaging on the $\sim10^{3}$ observations per object, each at the $\sim0.1-1$~mas level, and it relies on the self-calibration capability of the instrument, as well as on complementary photometric and spectroscopic data collected on board. The latter will also make it possible to include in the Gaia catalog a multiband spectro-photometric classification and the radial velocities of the objects brighter than $G\approx16.2$ \citep{2016A&A...595A...1G}.

\begin{figure}
	\begin{center}
		\includegraphics[width=\hsize]{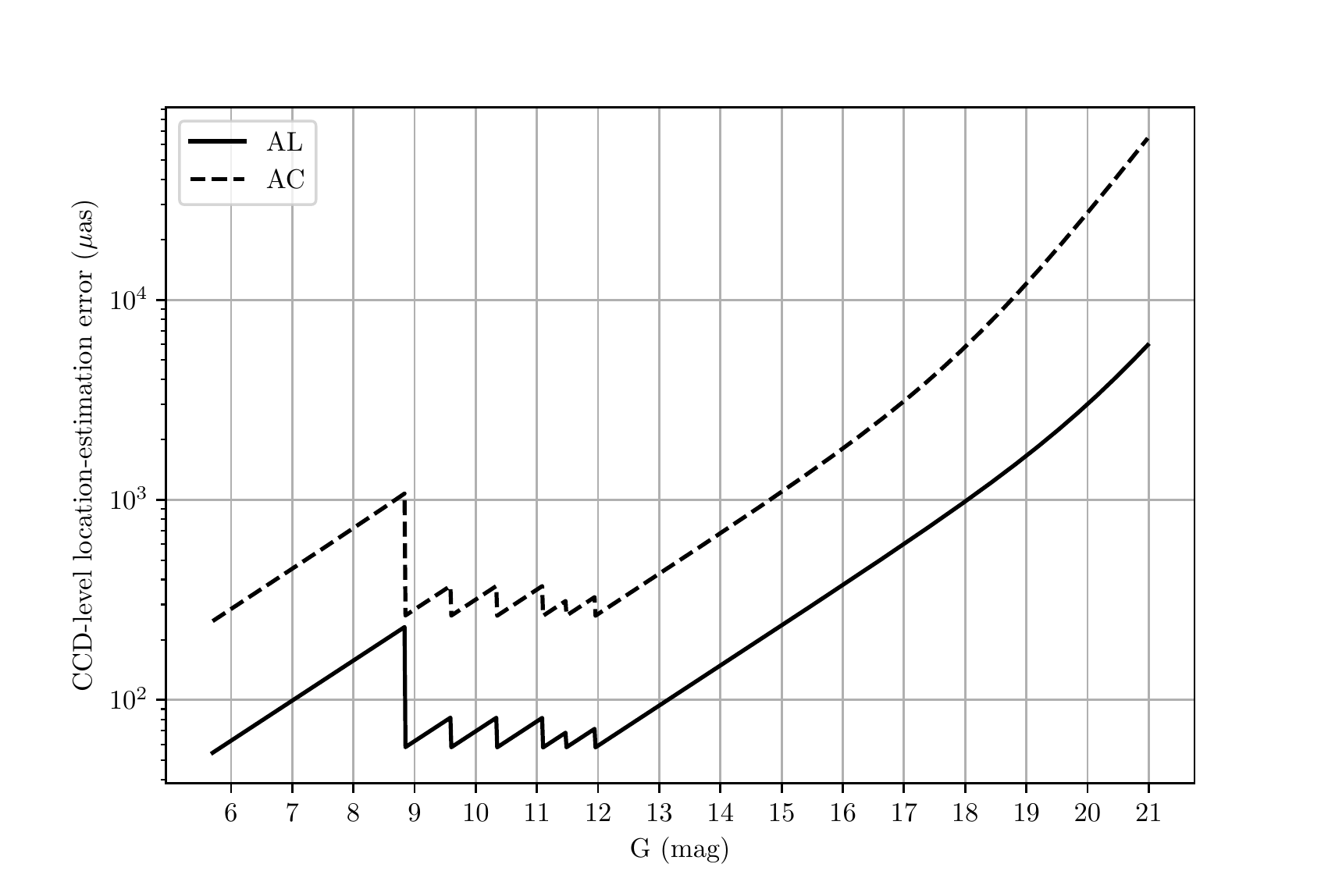}
		\caption{\label{fig:GaiaAccEst}Approximate estimation of the Gaia single-measurement along- and across-scan accuracy (AL and AC) as function of the $G$ magnitude. The oscillating shape up to mag 12 is due to the use of CCD gates in order to avoid saturation for bright objects. The estimation is the result of a best-fit model that averages over the whole Astrometric Field (AF).}
	\end{center}
\end{figure}

\begin{figure}
	\begin{center}
		\includegraphics[width=\hsize]{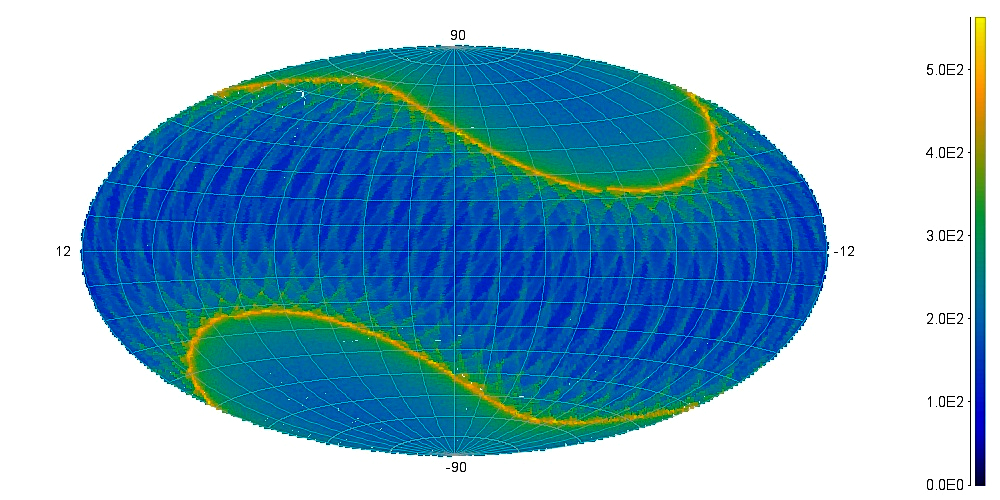}
		\caption{\label{fig:GaiaFreqObs}Frequency of observations as function of equatorial coordinates due to Gaia scanning law (blue $\approx 50$ - yellow $\approx 500$).}	
	\end{center}
\end{figure}

The sky is scanned according to a predetermined nominal scanning law (NSL) whose parameters are fixed at the beginning of the operational phase and actively controlled by the micro-thrusters of the on-board Attitude and Orbit Control Systems (AOCS). The NSL is characterized by constant spin and precession rates, which implies that objects cross the focal plane at quasi-constant speed. As a consequence, in principle the duration of a single observation is the same for any star, and therefore the single-measurement accuracy is strongly magnitude-dependent (Fig.~\ref{fig:GaiaAccEst}). Moreover, the scanning law parameters induce a complete scan of the celestial sphere every six months, but with a non-uniform coverage in terms of the number of single transits over a specific region, which implies that the number of observations of each object depends on its coordinates (Fig.~\ref{fig:GaiaFreqObs}). Therefore, the final astrometric accuracy of a specific object mainly depends on its magnitude and location on the sky.

Contrary to what usually happens for a space mission, for this mission ESA was in charge of the full development of the satellite, including the payload. The scientific community instead, organized in the Gaia Data Processing and Analysis Consortium (DPAC, \citealp{2007LL...FM03002}) funded by the national space agencies, has been in charge of the establishment of the data reduction pipelines of the mission. The software is developed by nine Coordination Units (CUs) while six Data Processing Centres (DPCs) spread all over Europe are commissioned of managing and running the pipelines.

The realization of the complete catalog is a complicated process in which the definition of the global reference frame is realized by a procedure called global astrometric sphere reconstruction. The latter operates with global astrometric data reduction techniques on a subset of $\sim10^{7}\text{--}10^{8}$ ``primary stars,'' similar in number to that of other modern global astrometric catalogs \citep{2013AJ....145...44Z,2015AJ....150..137Q} and mainly selected at the bright end of the Gaia objects. This allows to set the astrometric parameters of these objects with respect to a standard reference system called Barycentric Celestial Reference System (BCRS, \citealp{2005USNOC.179.....K}) thus providing the first materialization of such a frame of reference. The refinement of the instrument calibration needed to achieve the required catalog accuracy is carried out as part of the primary stars' processing. Once the reference system and the calibration parameters have been established, the measurements of the remaining stellar objects (secondary stars) can be reduced by considering only their astrometric parameters as unknowns. This allows to attach them to the primary stars and therefore to densify the reference frame. Basically, a star can be included in the primaries subset when its astrometric model can be described by the classical 5 parameters. Multiple stars with too short a period, or stars with a variability too large are examples of objects that cannot belong to the primaries. The two-steps process described above, namely the one including the global sphere reconstruction and the reduction of the secondary stars, is realized within the CU3, ``Core Processing'', by a pipeline called Astrometric Global Iterative Solution (AGIS; \citealt{2011ExA....31..215O,2012A&A...538A..78L}).


The global sphere reconstruction, and the materialization of a celestial reference system, has an absolute character (in the sense that it defines the reference system instead of giving the coordinates of the stars with respect to an already existing one) which implies an intrinsic difficulty at insuring the correctness of the astrometric parameters and at the same time the risk of propagating these errors everywhere in Astrophysics. This issue is well known to the scientific community, and significant effort is usually paid to both the tasks of internal verifications and of cross-checking comparison with different datasets (see, e.g.\ \citealt{2016A&A...595A...4L,2017A&A...599A..67C,2017ApJ...840L...1M,2018AAS...23143602F}). In the case of the Gaia forerunner, HIPPARCOS, the astrometric community provided the final catalog only after having compared two sphere reconstructions realized by two independent consortia. In the case of Gaia the task is so big that it is not feasible to repeat the same approach, but the DPAC decided to constitute an Astrometric Verification Unit (AVU) within the CU3 with the goal of replicating in an independent way three specific tasks of particular importance for the sphere reconstruction, namely the Astrometric Instrument Model (AIM; \citealt{2012SPIE.8449E..0FB, 2014SPIE.9150E..0KB}) for the instrument and focal plane calibration, the Basic Angle Monitoring (BAM; \citealt{2014RMxAC..45...35R,2019A&A...InPrep.R}) which has to determine the variations of the lines of sight of the double telescope, and the Global Sphere Reconstruction (GSR) whose aim is to provide an independent sphere reconstruction and to compare it with the AGIS one.

Since its goal is to allow a cross-checking of the sphere solution, in the sense of the reference system determination, GSR does not replicate the entire AGIS pipeline, but is limited to the processing of primary sources. Moreover, GSR depends on AGIS both for the determination of such sources -- even though it has the capability of providing an independent selection -- and for the rejection of time intervals with noisy attitude which must not enter in the sphere reconstruction. On the other hand, GSR is in charge of the comparison task between its own solution and that of the AGIS pipeline.

In this paper we adopt the following conventions and notations:
\begin{enumerate}
	\item in general, bold upright letters refers to 3D Euclidean vectors ($\boldsymbol{\mathrm{x}}$) while for basis unit vectors the notation $\mathbf{e}_{\hat{a}}$, $a=1,2,3$ is used;
	\item four-vectors are indicated by bold italic letters ($\boldsymbol{x}$) or in index notation with Greek indexes, that is $x^{\alpha}$, $\alpha=0,1,2,3$ with 0 referring to the time coordinate;
	\item the signature of the metric $g_{\alpha\beta}$ is $+2$;
    \item the symbol $\eta_{\alpha\beta}$ denotes the Minkowskian metric $\mathrm{diag}\{-1,1,1,1\}$;
	\item tetrad unit vectors are identified by $\boldsymbol{e}_{\hat{\alpha}}$, $\alpha=0,1,2,3$, while $\boldsymbol{e}_{\hat{a}}$ are the tetrad spatial axes;
	\item the proper time of an observer is indicated with the Greek letter $\tau$, while $t$ is the coordinate time;
	\item we adopt the symbol $\partial_{\alpha}$ as a shorthand notation for the derivative $\partial/\partial x^{\alpha}$.
\end{enumerate}

\begin{figure}
	\begin{center}
		\includegraphics[width=\hsize]{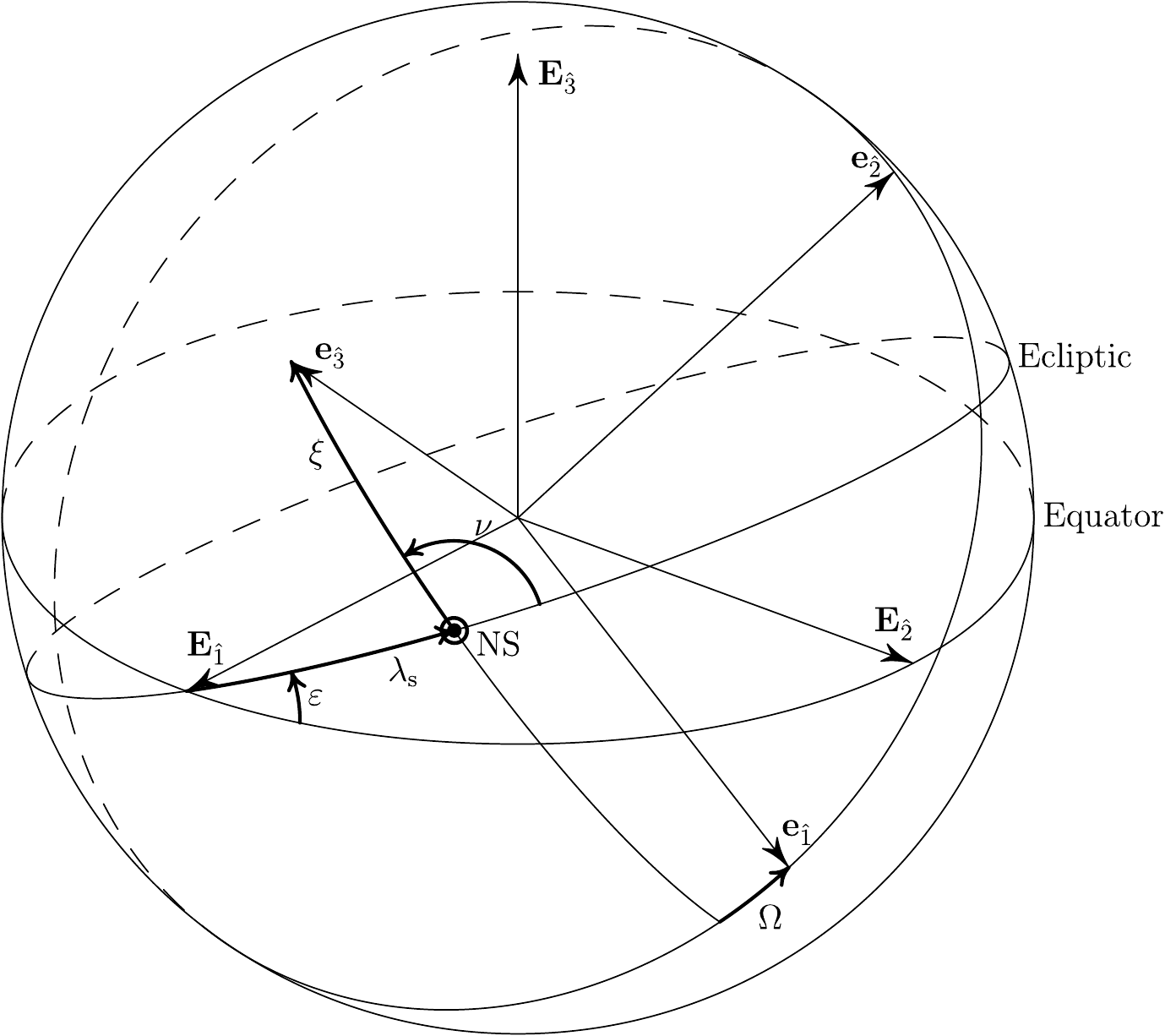}
		\caption{\label{fig:Gaia_NSL}Parametrization of the Gaia Nominal Scanning Law with respect to an equatorial reference system $\left\{\mathbf{E}_{\hat{1}},\mathbf{E}_{\hat{2}},\mathbf{E}_{\hat{3}}\right\}$. NS is the Nominal Sun.}	
	\end{center}
\end{figure}

\section{Modeling the observations for the global astrometric sphere reconstruction\label{sec:Modeling-the-observations}}

\subsection{Geometric characterization of the observable}
In a purely Euclidean geometric view, the NSL gives at each instant the orientation of the Satellite Reference System (SRS) that is of the reference triad $\left\{ \mathbf{e}_{\hat{a}}\right\} $, $a=1,2,3$ attached to the satellite, whose origin coincides with the barycentre of Gaia, with respect to the inertial reference system of the catalog. The position and orientation of the satellite results from the combination of three independent movements: the orbital motion of the satellite, which follows a Lissajous trajectory around the L2 point of the Sun-Earth system, the constant-rate satellite spin around the axis $\mathbf{e}_{\hat{3}}$, and the precession of this axis around the Sun-satellite direction, with constant rate and keeping a constant solar aspect angle $\xi$. Thus in an equatorial coordinate system the SRS is completely determined by five parameters: the heliotropic angles $\left(\xi,\nu,\Omega\right)$, the obliquity of the Ecliptic $\varepsilon$ and the longitude of the nominal Sun $\lambda_{\mathrm{s}}$ (Fig.~\ref{fig:Gaia_NSL}). Obviously the instantaneous orientation of the satellite depends also on the initial conditions of the NSL. Since the solar aspect angle $\xi$ is set to \ang{45}, the actual scanning law is fully determined by setting the remaining three degrees of freedom, namely $\nu$, $\Omega$ and $\lambda_{\mathrm{s}}$, at a given reference time.

\begin{figure}
	\begin{center}
		\includegraphics[width=0.92\hsize]{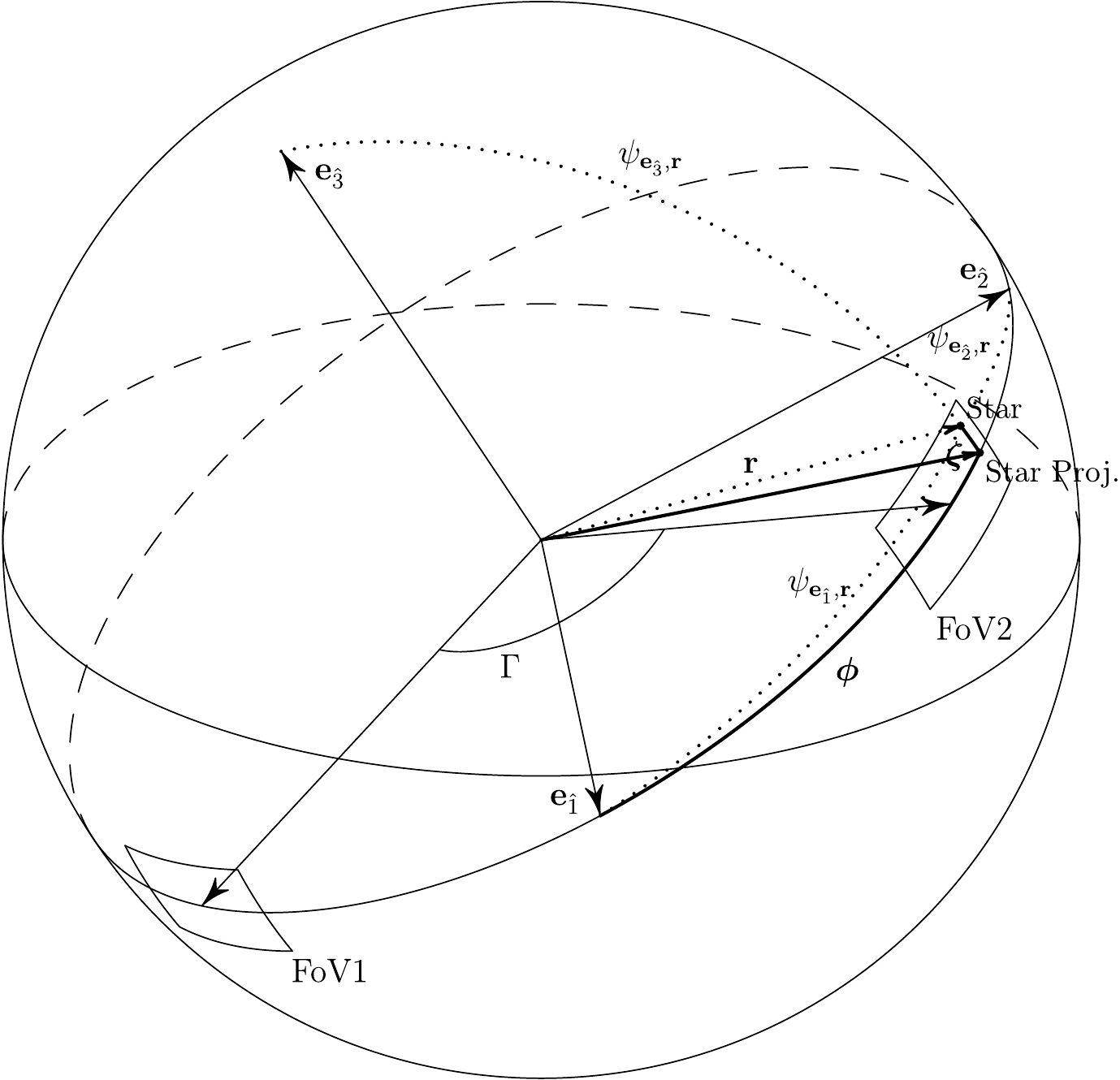}
		\caption{\label{fig:Gaia_measurements}Representation of the AL ($\phi$) and AC ($\zeta$) Gaia
        		measurements with respect to the satellite attitude $\left\{
                \mathbf{e}_{\hat{1}},\mathbf{e}_{\hat{2}},\mathbf{e}_{\hat{3}}\right\} $.}	
	\end{center}
\end{figure}

In this geometrical model the two Fields-of-View (FoVs) of Gaia lie on the satellite' scanning plane $\mathbf{e}_{\hat{1}}-\mathbf{e}_{\hat{2}}$ and are pointing symmetrically with respect to $\mathbf{e}_{\hat{1}}$, separated by an (approximately) constant \emph{Basic Angle }(BA) $\Gamma$. If $\mathbf{r}$ is the position vector of a point-like source, in the SRS the basic observables of Gaia can be represented by its along-scan measurement (AL) that is the abscissa $\phi$, which is the angle between the projection of $\mathbf{r}$ on the instantaneous scanning plane and $\mathbf{e}_{\hat{1}}$, and the across-scan measurement (AC) $\zeta$ (Fig.~\ref{fig:Gaia_measurements}).

These quantities can be represented in terms of the direction cosines of $\mathbf{r}$ with respect to the SRS
\begin{align}
	\cos\phi & =\frac{\cos\psi_{\left(\hat{1},\mathbf{r}\right)}}
    				{\sqrt{1-\cos^{2}\psi_{\left(\hat{3},\mathbf{r}\right)}}}
                    \label{eq:cosphi}\\
	\sin\zeta & =\cos\psi_{\left(\hat{3},\mathbf{r}\right)},\label{eq:sinzeta}
\end{align}
while the direction cosine of $\mathbf{r}$ with respect to the axis
$\mathbf{e}_{\hat{a}}$ is
\begin{equation}
	\cos\psi_{\left(\hat{a},\mathbf{r}\right)}=\frac{\mathbf{e}_{\hat{a}}\cdot\mathbf{r}}
    		{\left(\mathbf{r}\cdot\mathbf{r}\right)^{1/2}}.\label{eq:dircos_eucl}
\end{equation}
It is worth stressing that these modeling equations leave a potential ambiguity in the sign of the abscissa $\phi$, which should be negative when the observation lies in FoV1 and positive in FoV2. However, the FoV of each observation is specified in the Gaia data, which removes this ambiguity.

\subsection{Unknown parameters in the Gaia sphere reconstruction}
In general $\mathbf{e}_{\hat{a}}$ and $\mathbf{r}$ will be functions of the time of observation, of the attitude parameters and of the source parameters respectively. The latter are the classical parallax, positions and proper motions $\left(\varpi,\alpha,\delta,\mu_{\alpha},\mu_{\delta}\right)$ because in this case all the sources are primaries, while the attitude parameters have to be counted among the unknowns too, because they cannot be determined independently at the level needed for the Gaia accuracy. The attitude parameters and the explicit form of the functions, however, will depend on the specific astrometric model adopted to represent the observation.

Gaia is also a self-calibrating mission, in the sense that there are no metrologic instruments, with the only exception of the BAM, dedicated to the calibration. All the calibrations except for the monitoring of the Basic Angle variations are done using the Gaia observations only. This task is accomplished in two different ways. One is a daily calibration done by the two independent pipelines, First Look (FL) and, in the AVU subsystem, AIM; the other one is embedded in the sphere solution. The daily calibration is needed to monitor the instrument, using science data to trace directly the instrument response thanks to the repeated measurements of stars over the field, but as long as this procedure relies on a daily approach, it cannot reach the accuracy needed by the sphere reconstruction. Therefore, as it will be better specified in Section~\ref{subsec:Instrument-parameters}, AGIS and GSR have to introduce appropriate calibration parameters as additional unknowns of their models.

Finally, there exist also another class of unknowns called global parameters. Such name derives from their presence in all the rows of the system of equations that is built from the Gaia observations, as it will be explained in Section~\ref{sec:LinEqSyst}, and typically their number is the smallest with respect to all the other classes.

By indicating with $\mathbf{x}^{\mathrm{S}}$, $\mathbf{x}^{\mathrm{A}}$, $\mathbf{x}^{\mathrm{C}}$ and $\mathbf{x}^{\mathrm{G}}$ the list of source, attitude, calibration and global parameters respectively, Eqs.~(\ref{eq:cosphi}) and (\ref{eq:sinzeta}) can be formally written as
\begin{align}
	\cos\phi & =f_{\phi}\left(\mathbf{x}^{\mathrm{S}},\mathbf{x}^{\mathrm{A}},
 							  \mathbf{x}^{\mathrm{C}},\mathbf{x}^{\mathrm{G}}
                        \right)\label{eq:cosphi_gen}\\
	\sin\zeta & =f_{\zeta}\left(\mathbf{x}^{\mathrm{S}},\mathbf{x}^{\mathrm{A}},
    							\mathbf{x}^{\mathrm{C}},\mathbf{x}^{\mathrm{G}}
                          \right).\label{eq:sinz_gen}
\end{align}

\subsection{Gaia accuracy and the relativistic observable\label{sec:rel-obs}}
The target accuracy of Gaia is at the sub-mas level at least, moreover  the catalog will be released in the BCRS and linked to the International  Celestial Reference Frame (ICRF) defined by VLBI observations \citep{2015AJ....150...58F}.\footnote{The ICRF is the so-called realization of the International Celestial Reference System (ICRS) \citep{2006ITN....34....1S}. Likewise, the Gaia catalog is the realization of the BCRS.} For these reasons it is necessary to use a mathematical model of the observations based on General Relativity. Basically, the task of this model is to provide a relativistically consistent formula to replace Eq.~(\ref{eq:dircos_eucl}). This requires the relativistic equivalent of the photon's incoming direction (\textbf{r}), of the satellite attitude (the SRS), and of the definition of scalar product.

In General Relativity the trajectory of a light ray connecting a source with the observer is defined by a set of four-dimensional events $x^{\alpha}\left(s\right)$, for which the condition  $g_{\alpha\beta}\mathrm{d}x^{\alpha}\mathrm{d}x^{\beta}=0$ holds, and is thus called null geodesic, parametrized by a generic affine parameter $s$. This trajectory can be found by integrating the geodesic equations
\begin{equation}
	\frac{\mathrm{d}^{2}x^{\alpha}}{\mathrm{d}s^{2}}+\Gamma_{\mu\nu}^{\alpha}\frac{\mathrm{d}x^{\mu}}
    	{\mathrm{d}s}\frac{\mathrm{d}x^{\nu}}{\mathrm{d}s}=0,\label{eq:geod_eqs}
\end{equation}
where
\begin{equation}
	\Gamma_{\mu\nu}^{\alpha}=\frac{1}{2}g^{\alpha\rho}
    	\left(\partial_{\mu}g_{\nu\rho}+\partial_{\nu}g_{\rho\mu}-\partial_{\rho}g_{\mu\nu}\right),
\end{equation}
and the role of \textbf{r }is usually played by the value of the tangent to the null geodesic at the point of observation, namely by the four-vector $k^{\alpha}=\mathrm{d}x^{\alpha}/\mathrm{d}s$ evaluated at that point. The photon's incoming direction computed in this way obviously depends on the relativistic effects induced by the metric and thus by the gravitational effects of the massive bodies, such as, for example, the so-called light deflection.

Any measurement is always defined with respect to a specific observer, identified with its four-velocity $u^{\alpha}=\mathrm{d}x^{\alpha}/\mathrm{d}\tau$, which induces a natural 3+1 splitting of the spacetime, that is the identification of the space and of the time respectively associated to such observer. Specifically, $u^{\alpha}$ is tangent to the time axis of the observer, represented by its worldline, and the 3D space is the subspace orthogonal to $u^{\alpha}$. Any spatial measurement, therefore, is performed by means of the operator $h_{\alpha\beta}=g_{\alpha\beta}+u_{\alpha}u_{\beta}$ which can project the appropriate four-dimensional quantities onto this 3D subspace, and is thus used as a replacement of the usual 3D dot product. For example, if in Euclidean geometry the angle $\alpha$ between two vectors $\mathbf{r}_{1}$ and $\mathbf{r}_{2}$ satisfies the relation $\cos\alpha=\mathbf{r}_{1}\cdot\mathbf{r}_{2}/\left(r_{1}r_{2}\right)$, with $r=\left(\mathbf{r}\cdot\mathbf{r}\right)^{1/2}$, the corresponding general relativistic expression will be
\begin{equation}
\cos\alpha=\frac{h_{\alpha\beta}k_{1}^{\alpha}k_{2}^{\beta}}{\left(h_{\mu\nu}k_{1}^{\mu}k_{1}^{\nu}\right)^{1/2}\left(h_{\rho\sigma}k_{2}^{\rho}k_{2}^{\sigma}\right)^{1/2}}.
\end{equation}
It is worth noticing that in this way the aberration effects are automatically accounted for by the inclusion of the observer's four-velocity in $h_{\alpha\beta}$.

Finally, one standard way to handle the SRS is based on the so-called tetrad formalism \citep{2010cmcs.book.....D}, which is based on the possibility of identifying a locally Lorentzian reference system associated to a specific observer $u_{\mathrm{s}}^{\alpha}$, that is a set of four-dimensional axes $\left\{ \boldsymbol{e}_{\hat{\alpha}}\right\} $, $\alpha=0,\ldots,3$ called tetrad defined by the conditions
\begin{equation}
	g_{\mu\nu}e_{\hat{\alpha}}^{\mu}e_{\hat{\beta}}^{\nu}=\eta_{\alpha\beta},\quad e_{\hat{0}}^{\alpha}
    \equiv u_{\mathrm{s}}^{\alpha}.\label{eq:tetrad_cond}
\end{equation}
In this way, we are inducing the same 3+1 splitting of above with the addition of a set of three four-dimensional axes $\left\{ \boldsymbol{e}_{\hat{a}}\right\} $, $a=1,2,3$ belonging to the 3D space of $u^{\alpha}$. Since by definition $g_{\mu\nu}u^{\mu}e_{\hat{a}}^{\nu}=g_{\mu\nu}e_{\hat{0}}^{\mu}e_{\hat{a}}^{\nu}=0$ and $g_{\mu\nu}e_{\hat{a}}^{\mu}e_{\hat{b}}^{\nu}=\eta_{ab}=\delta_{ab}$, these axes are spatial (in the sense that they are orthogonal to $u_{\mathrm{s}}^{\alpha}$ and thus lie on its 3D space) and orthonormal, so they provide the replacement for the Euclidean triad of Eq.~(\ref{eq:dircos_eucl}) and can be used to define the attitude (SRS) of the satellite. The orientation of the spatial axes, in fact, is constrained only by the orthonormality condition, which implies that any spatial 3D rotation brings to a new tetrad with a different spatially oriented triad.

By combining all the above considerations we can write the general relativistic expression for the direction cosine with respect to a general spatial axis of the tetrad as
\begin{equation}
	\cos\psi_{\left(\hat{a},\boldsymbol{k}\right)}=\frac{h_{\alpha\beta}e_{\hat{a}}^{\alpha}k^{\beta}}
    		{\left(h_{\mu\nu}k^{\mu}k^{\nu}\right)^{1/2}}.\label{eq:dircos_rel}
\end{equation}
On the other hand, the AL and AC measurements of Eqs.~(\ref{eq:cosphi}) and (\ref{eq:sinzeta}) are defined in the 3D subspace of $u_{\mathrm{s}}^{\alpha}$ as long as the direction cosines are the relativistic ones, therefore they do not need any relativistic replacement.

We can thus summarize the general procedure applied to build the relativistic model of the Gaia observations used in GSR with the following list:
\begin{enumerate}
	\item integrate the geodesic equations (\ref{eq:geod_eqs}) for the given metric (the BCRS one, in the case of Gaia) and find an expression of $k^{\alpha}$ at the point of observation as function of the source's
coordinates;
	\item find the appropriate expression $u_{\mathrm{s}}^{\alpha}$ of the barycentric motion of Gaia in the BCRS;
	\item use the above four-velocity and the BCRS metric to compute the spatial projector $h_{\alpha\beta}$;
	\item find the Gaia relativistic tetrad by a two-steps procedure:
	\begin{enumerate}
		\item use Eqs.~(\ref{eq:tetrad_cond}) to define a local tetrad, whose origin is comoving with the barycentre of Gaia and whose spatial axes are kinematically parallel to those of the BCRS (``boosted tetrad'');
		\item 3D rotate the spatial axes to make them coincide with the satellite orientation, thus realizing a tetrad associated to the Gaia barycentric and attitude motion $\left\{ \boldsymbol{e}_{\hat{0}},\boldsymbol{e}_{\hat{a}}\right\} $;
	\end{enumerate}
	\item use $k^{\alpha}$, $h_{\alpha\beta}$ and $e_{\hat{a}}^{\alpha}$ in Eq.~(\ref{eq:dircos_rel}) to compute the needed direction cosines;
	\item use Eqs.~(\ref{eq:cosphi}) and (\ref{eq:sinzeta}) to compute the Gaia measurements.
\end{enumerate}
Within this general framework, the accuracy needed for the astrometric model is set by that of the Gaia measurements and of its final catalog. In practice it is useful to link the two by means of the Post-Newtonian ``bookkeeping'' \citep{1993tegp.book.....W}. In this way, considering that the typical velocities in the Solar System are $\simeq30\:\mathrm{km}/\mathrm{s}$, the so-called 1PN order of $\left(v/c\right)^{2}$ would correspond to a $\sim10^{-8}\:\mathrm{rad}$ in angular accuracy, that is to the mas level, allowing one to set the befitting model accuracy at the 1.5PN order $\left(v/c\right)^{3}$, corresponding to $10^{-12}\:\mathrm{rad}$, or $\sim0.1\:\mu$as.

Actually, this has to be considered just a first, although convenient, guess. Things can be much different in the real case, and the final accuracy strongly depends on other factors, like the geometry of the observations or the observation strategy. For example, in a global problem like the sphere solution, for large regions of the sky one can safely rely on less stringent accuracy requirements, whereas relative astrometric observations close to Solar System planets would surely need a more accurate model. In this context, we also stress that Gaia-level final accuracies have also been reached in the context of differential astrometry \citep{2018ApJ...856L...6B}.

GSR is designed to be flexible in regard to the relativistic modeling of the observable, and different relativistic models are under development to add value to the verification purposes of this pipeline. We are dealing specifically with this issue in a forthcoming publication (\citealt{2019Models...InPrep..V}, in preparation) whereas in the next subsection we are concentrating on the implementation of GSR2, namely the present version of our pipeline.

\subsection{Integration of the geodesic equation in the current pipeline implementation\label{sec:int_geod}}
The geodesic equations are integrated in the single-body PPN-Schwarzschild metric, where the tangent to the null geodesic $k^{\alpha}$ is a function of three constants of motion $E_{*}$, $\Lambda_{*}$, and $\lambda_{*}$ characteristic of the geodesic connecting the observer with the observed object in the Schwarzschild metric,\footnote{$E_{*}$, $\Lambda_{*}$, and $\lambda_{*}$ have the meaning of the photon's energy, total angular momentum over energy and $z$-angular momentum over energy.} of the satellite position $\left(r_{\mathrm{s}},\theta_{\mathrm{s}},\phi_{\mathrm{s}}\right)$, and of the PPN-$\gamma$ parameter:
\begin{equation}
	k^{\alpha}=k^{\alpha}\left(E_{*},\Lambda_{*},\lambda_{*},r_{\mathrm{s}},
    	\theta_{\mathrm{s}},\phi_{\mathrm{s}},\gamma\right).\label{eq:k-cost-mot}
\end{equation}

First of all, it is worth stressing that the accuracy of the Gaia measurement allows for the estimation of the $\gamma$ parameter as a by-product of the global sphere reconstruction \citep{2003A&A...399..337V}. This implies that this parameter can be treated as an unknown belonging to the vector of global parameters $\mathbf{x}^{\mathrm{G}}$.

Moreover, the constants of motion can be written as functions of the astrometric unknowns at the time of observation, by means of the same principle used in \citet{2003A&A...399..337V} and references therein, which in summary eliminates the dependence on $E_{*}$ and provides two equations that implicitly relate the stellar position with $\Lambda_{*}$ and $\lambda_{*}$
\begin{equation}
	\begin{array}{rcl}
	f_{r}\left(\varpi_{*},r_{\mathrm{s}},\Lambda_{*},\gamma\right) & = & 		
    		f_{\theta}\left(\theta_{*},\theta_{\mathrm{s}},\lambda_{*}/\Lambda_{*}\right)\\
	\bar{f}_{\theta}\left(\theta_{*},\theta_{\mathrm{s}},\Lambda_{*},\lambda_{*}\right) & = & 
    		f_{\phi}\left(\phi_{*}-\phi_{\mathrm{s}}\right)
	\end{array}\label{eq:syst-algebraic}
\end{equation}
where $\varpi=a/r_{*}$ is the parallax, $a=1\:\mathrm{AU}$ is the parallaxes baseline, $\theta_{*}=\pi/2-\delta$ and $\phi_{*}\equiv\alpha$. Finally, the proper motion can be easily included in the model simply by considering the integration limits on $\theta_{*}$ and $\phi_{*}$ as function of the (coordinate) time, that is
\begin{eqnarray}
	\delta & \equiv & \delta(t)=\delta(t_0)+\mu_{\delta}\times(t-t_{0})\label{eq:mu-theta}\\
	\alpha & \equiv & \alpha(t)=\alpha(t_0)+\mu_{\alpha}\times(t-t_{0})\label{eq:mu-phi}
\end{eqnarray}
where $t_{0}$ is the epoch of the astrometric catalog or an equivalent reference time. The above formulae neglect the effect on spherical coordinates at second order in proper motions (see e.g. \citet{1985spas.book.....G}, p.264), which are negligible in most cases. For example, in the catalog of more than 900,000 simulated stars considered for the demonstration run of this paper, this effect is larger than $5~\mu$as after 5 years for just 22 objects in $\alpha$ and 5 objects in $\delta$.

It is clear from this summary that, as long as the metric is a Schwarzschild one due to the gravitational pull of the Sun, the accuracy of the null geodesic integration is not the required one. At the same time it is obvious that the Gaia accuracy can be attained when the observing direction is sufficiently far from the gravitational perturbation of the Solar System bodies. One can approximately estimate these ``avoidance zones'' by comparing the estimated measurement accuracy with the Schwarzschild contribution to the light deflection of each single body as a function of its angular distance $\psi$ from the source \citep{1973grav.book.....M}
\begin{equation}
	\delta\psi=\frac{\left(1+\gamma\right)M}{c^{2}r_{\mathrm{s}}}
    	\sqrt{\frac{1+\cos\psi}{1-	\cos\psi}},\label{eq:Schw_defl_misner}
\end{equation}
where $M$ is the mass of the body and $r_{\mathrm{s}}$ its distance from the satellite. For example, the minimum angular distance that a source must have from Jupiter in order to keep the model accuracy below the $\sim10\:\mu$as, level is about 10.5 degrees.

On the other hand, the model accuracy can be enhanced by adding the estimated light deflection effects of Eq.~(\ref{eq:Schw_defl_misner}) to the pure Schwarzschild model; such contribution to the light deflection has to be computed at an appropriate retarded time which takes into account the finite speed of the light. These corrections are done separately for $\phi$ and $\zeta$ by projecting the $\delta\psi$ on the AL and AC direction according to the geometrical configuration of the observation. Clearly this is in no way an exact solution, but, as shown in (\citealt{2019Models...InPrep..V}, in preparation) where a more detailed description of the model will also be provided, numerical tests have proven that the accuracy of this model is much better than that of the pure Schwarzschild one.

\subsection{Satellite barycentric motion and attitude model}
As shown in \citet{2010A&A...509A..37C}, the four-velocity of the satellite can be written as $u_{\mathrm{s}}^{\alpha}=u_{\mathrm{s}}^{0}\left(\delta_{0}^{\alpha}+v_{\mathrm{s}}^{i}\delta_{i}^{\alpha}\right)\equiv u_{\mathrm{s}}^{0}\left\{ 1,v_{\mathrm{s}}^{x},v_{\mathrm{s}}^{y},v_{\mathrm{s}}^{z}\right\} $ where $v_{\mathrm{s}}$ is the coordinate velocity of the satellite and, using the IAU resolutions \citep{2005USNOC.179.....K},
\begin{equation}
	u_{\mathrm{s}}^{0}=1+\frac{U}{c^{2}}+\frac{1}{2}
    \frac{v_{\mathrm{s}}^{2}}{c^{2}}+\mathcal{O}\left(\frac{v^{4}}{c^{4}}\right).
\end{equation}
Here $v_{\mathrm{s}}^{2}=\left(v_{\mathrm{s}}^{x}\right)^{2}+\left(v_{\mathrm{s}}^{y}\right)^{2}+\left(v_{\mathrm{s}}^{z}\right)^{2}$ and $U$ is the (BCRS) $N$-body gravitational potential of the Solar System at the Gaia location, which implies that the four-velocity, contrary to the null geodesic, is expressed to the right accuracy.

The construction of the attitude tetrad, as anticipated in Eq.~(\ref{eq:tetrad_cond}), starts from this four-vector by setting $e_{\hat{0}}^{\alpha}=u_{\mathrm{s}}^{\alpha}$. Then one has to fix the spatial triad by considering the following transformations \citep{2010A&A...509A..37C,2011LL...LB00101}:
\begin{enumerate}
	\item the origin of the tetrad is located at the BCRS coordinates of the satellite barycentre, which induces a first transformation to the so-called local BCRS tetrad \citep{2003CQGra..20.4695B}. This is a tetrad at rest with respect to the BCRS, but whose spatial axes have a general relativistic contribution caused by the gravitational potential of the Solar System at that point;
	\item the satellite barycentre moves with four-velocity $u_{\mathrm{s}}^{\alpha}$ in the BCRS, which induces a second transformation to the boosted tetrad with spatial axes $\left\{ \boldsymbol{e}_{\hat{a},\mathrm{b}}\right\}$ whose origin is comoving with the barycentre of the satellite.\footnote{This tetrad is thus the CoMRS of the Gaia DPAC nomenclature \citep{2012A&A...538A..78L}.} If the four-velocity of the BCRS, and therefore also that of the local BCRS, is $u^{\alpha}$, then at each instant the comoving tetrad can be computed with a special relativistic transformation, namely a boost, on the previous one whose Lorentz factor is $\gamma_{\mathrm{L}}=\left(1-v_{\mathrm{s}}^{2}/c^{2}\right)^{-1/2}=-u_{\mathrm{s}}^{\alpha}u_{\alpha}$;
	\item the satellite orientation, that is the SRS $\left\{ \boldsymbol{e}_{\hat{a}}\right\} $, can be finally obtained by a Euclidean transformation of its spatial axes, $e_{\hat{a}}^{\alpha}=Re_{\hat{a},\mathrm{b}}^{\alpha}$, where $R$ is simply a 3D rotation matrix obtained by a standard attitude parametrization. AGIS uses quaternions, while GSR is based on the Modified Rodrigues Parameters (MRP) $\sigma_{1}$, $\sigma_{2}$ and $\sigma_{3}$ described, for example, in \citet{1996JAS...44...1S}. Since MRP representation uses only three parameters instead of the four of the quaternions, adopting the MRP representation allows to reduce the number of attitude unknowns. Moreover, quaternions require the enforcing of a normalization condition, implemented through additional constraint equations, due to their redundant parametrization. On the other hand, such redundancy allows to avoid the problem of singularities that is present in other parametrizations (including the MRP). The scanning law of Gaia, however, is always sufficiently far from the singularity points of the MRP.
\end{enumerate}
Despite their different accuracy, the null geodesic and the attitude parts stem from similar mathematical assumptions, which makes them mutually compatible at the $\left(v/c\right)^{2}$ order. Basically, the main requirement is that the metric has the form
\begin{equation}
	g_{\alpha\beta}=\eta_{\alpha\beta}+h_{\alpha\beta}+{\cal O}\left(h^{2}\right)\label{eq:pert-metric}
\end{equation}
where, for the null geodesic integration, we retained only the $\left(v/c\right)^{2}$-order perturbation term given by the gravitational potential of a massive body $h_{00}^{(2)}=2U/c^{2}$, so that $h_{00}\sim h_{ij}=h_{00}^{(2)}\delta_{ij}+{\cal O}\left(v^{4}/c^{4}\right)$, while the attitude model can include also higher order terms, namely $h_{0i}\sim v^{3}/c^{3}+{\cal O}\left(v^{5}/c^{5}\right)$ which enter in the definition of the local BCRS tetrad.

Considering the problem of the attitude representation, it has to be recalled that, regardless of the specific parametrization chosen to represent the attitude matrix, each component of such parametrization has to be expressed with a finite and time-dependent number of attitude unknowns $\mathbf{x}^{\mathrm{A}}$. The required final accuracy of the measurements does not permit the use of a physical model for the attitude due to several factors; for example, the micro-propulsion system introduces a high-frequency noise. As in AGIS, then, an attitude parameter (that is a quaternion or an MRP component) is represented by a purely numerical expansion written as a linear combination of time-dependent polynomial functions $B\left(t\right)$ of degree $M-1$ called B-Splines \citep{Ahlberg1967} whose characteristics are summarized in the following.

The expansion of the $j$-th component of a generic representation, say $S$, reads
\begin{equation}
	S_{j}\left(t\right)=\sum_{n=0}^{N-1}c_{n}^{\left(j\right)}B_{n}\left(t\right),\label{eq:MRP_bspline}
\end{equation}
where $c_{n}^{\left(j\right)}$ are unknown attitude coefficients to be determined. The function $S_{j}\left(t\right)$ is defined in a time interval $\left[t_{\mathrm{beg}},\ldots,t_{\mathrm{end}}\right]$ divided in $K>0$ sub-intervals identified by a sequence of instants $\left\{ \tau_{k}\right\} $, $k=0,\ldots,K$ called nodes, which constitutes the so-called support of the series. The B-Splines are useful in our case because their support is minimal, that is $B_{n}\left(t\right)\neq0$ only in $M$ sub-intervals, so if $\tau_{n}<t<\tau_{n+1}$ then,
\begin{equation}
	S_{j}\left(t\right)=f\left(c_{n-M/2}^{\left(j\right)},\ldots,
    	c_{n-2}^{\left(j\right)},c_{n-1}^{\left(j\right)},c_{n}^{\left(j\right)},
        c_{n+1}^{\left(j\right)},\ldots,c_{n+M/2-1}^{\left(j\right)}\right),
\end{equation}
in the sense that at each time the expansion depends only on $M$ unknowns. For Gaia the expansion is in cubic B-splines, that is\ $M=4$, and therefore
\begin{equation}
	S_{j}\left(t\right)=f\left(c_{n-2}^{\left(j\right)},c_{n-1}^{\left(j\right)},
    	c_{n}^{\left(j\right)},c_{n+1}^{\left(j\right)}\right).
\end{equation}
Finally, having split a time segment in $K$-1 intervals, the resulting number of degrees of freedom, and thus of attitude unknowns, for that segment is given by $N=K+M-1$.

Another important point in using a series expansion like the B-Splines lies in the possibility of exploiting the linearization of the observation equations to implement the so-called differential attitude approach. As for the integration of the geodesics, this will be formulated in more detail in a forthcoming paper, but since the motivation and the principle of this approach is directly linked to the solution method, Section~\ref{sec:Reconstructing} will present a line-of-principle description from the mathematical point of view.

\subsection{Instrument parameters\label{subsec:Instrument-parameters}}
As anticipated in the general overview of Section~\ref{sec:Modeling-the-observations}, except for the BAM, Gaia has no onboard metrologic instrument, and a set of long-term calibration parameters $\mathbf{x}^{\mathrm{C}}$ has to be introduced among the unknowns of the global sphere reconstruction. The instrument calibration parameters currently used by GSR are those described in \citet{2012A&A...538A..78L}. In this model, like for the attitude, the calibration is done by a purely numerical description of the perturbations of the two field angles $\eta$ and $\zeta$ on the focal plane, respectively in the along- and across-scan direction. The parameters of this model can be divided in two classes: geometric and spectrophotometric. In the former the deviations $\mathrm{d}\eta$ and $\mathrm{d}\zeta$ can be described by geometrical deviations at the level of a single CCD of the astrometric focal plane from the nominal configurations, namely shifts, shears plus rotations and distortions, while in the latter magnitude- and spectrum- dependent shifts $\mathrm{d}\eta$ are introduced.

In practice the AL measurement $\phi$ in Eq.~(\ref{eq:cosphi_gen}) is defined as
\begin{equation}
	\phi=\eta\pm\frac{\Gamma}{2},
\end{equation}
where $\Gamma$ is the Basic Angle value (BA) and the sign is positive when the observation is on the preceeding Field of View (FoV2) and negative for the following Field of View (FoV1). The calibration model then puts
\begin{equation}
	\eta=\eta^{0}+\sum_{r=0}^{2}\Delta\eta_{rfnj}L_{r}^{*}
		\left(\tilde{\mu}\right)+\delta\eta_{nm}+C_{n}^{\mathrm{mag}}\left(G-G_{\mathrm{ref}}\right)
        +C_{fn}^{\mathrm{sp}}\left(\nu-\nu_{\mathrm{ref}}\right)\label{eq:eta_obs}
\end{equation}
in which $\eta^{0}$ is the nominal AL field angle and $\Delta\eta_{rfnj}$, $\delta\eta_{nm}$, $C_{n}^{\mathrm{mag}}$ and $C_{fn}^{\mathrm{sp}}$ are the above mentioned calibration unknowns, namely the AL large and small scale geometric corrections, and the magnitude- and spectrum- dependent shifts. The indices indicate the dependencies of these parameters with respect to the instrument configuration.

The large-scale AL parameters $\Delta\eta_{rfnj}$ depend on the FoV ($f$), the CCD index ($n$) and have a temporal variation so that there exists a different set of parameters extending over a certain time interval, indexed by $j$. Moreover, the shift, shear plus rotation, and distortion contributions are modeled by different orders $0\le r\le2$ of the Legendre polynomials $L_{r}^{*}\left(\tilde{\mu}\right)$ computed at the normalized AC pixel coordinate
\begin{equation}
	\tilde{\mu}=\frac{\mu-\mu_{0}+0.5}{1966},
\end{equation}
where $14\le\mu\le1979$ is the pixel coordinate of the measurement and $\mu_{0}=14$ is the pixel coordinate of the beginning of the light sensitive area of the CCD.

The small scale AL parameters $\delta\eta_{nm}$ depend on the CCD index $n$ and on the AC pixel index $m$, which goes from $1+\mu_{0}$ to $\mathrm{pixcols}+\mu_{0}$, where pixcols is the total number of illuminated pixels. The same indexes are used to identify the dependencies of the magnitude- and spectrum- dependent shifts, which need a reference magnitude and frequency $G_{\mathrm{ref}}$ and $\nu_{\mathrm{ref}}$ respectively.

Similarly, the AC measurement $\zeta$ of Eq.~(\ref{eq:sinz_gen}) is
\begin{equation}
	\zeta=\zeta^{0}+\sum_{r=0}^{2}\Delta\zeta_{rfnj}L_{r}^{*}
    	\left(\tilde{\mu}\right)+\delta\zeta_{nm}\label{eq:zeta_obs}
\end{equation}
where it is made evident that in the AC direction only the geometric parameters are taken into account. This calibration model introduces a degeneracy among the geometric parameters and the attitude which has to be removed by means of a set of appropriate constraint equations.

\section{Reconstructing the global astrometric sphere\label{sec:Reconstructing}}
\subsection{The linearized system of equations\label{sec:LinEqSyst}}
The global astrometric sphere reconstruction, in the case of Gaia, requires the solution of a large, sparse and overdetermined system of linearized equations in the least-squares sense. Each observation in fact is represented by a formula either as Eq.~(\ref{eq:cosphi}) or (\ref{eq:sinzeta}), which ultimately are functions of four types of unknowns, namely $\mathbf{x}^{\mathrm{S}}$, $\mathbf{x}^{\mathrm{A}}$, $\mathbf{x}^{\mathrm{C}}$, $\mathbf{x}^{\mathrm{G}}$, and together these observations produce a system of equations whose dimension depend on the number of unknowns $n_{\mathrm{unk}}$ and of observations $n_{\mathrm{obs}}$.

In the case of Gaia the total number of unknowns is $n_{\mathrm{unk}}=n^{\mathrm{S}}+n^{\mathrm{A}}+n^{\mathrm{C}}+n^{\mathrm{G}}$, where the number of unknowns for the sources is $n^{\mathrm{S}}=5n^{*}$, $n^{*}\sim10^{8}$ is the number of primary stars and $n^{\mathrm{A}}$, $n^{\mathrm{C}}$ and $n^{\mathrm{G}}$ are the total number of attitude, instrument and global unknowns respectively. Since generally $n^{\mathrm{G}}\ll n^{\mathrm{C}}\ll n^{\mathrm{A}}\ll n^{*}$, it is reasonable to put $n_{\mathrm{unk}}\simeq6n^{*}$. Moreover, we know that on average each star is observed about $10^{3}$ times during the mission lifetime, so the total number of observations is $n_{\mathrm{obs}}\sim10^{11}$.

Since $n_{\mathrm{obs}}\gg n_{\mathrm{unk}}$ the system is not only large but also overdetermined, so in principle it can be solved in the least squares sense, providing not only an estimation of the unknowns but also of their errors and correlations. Furthermore, each equation represents the observation of a single source in a specific time interval, therefore it depends only on (up to) 5 source unknowns and a comparably limited number of attitude and instrument parameters, which implies the sparseness of the system.

Finally, both Eqs.~(\ref{eq:cosphi}) and (\ref{eq:sinzeta}) are highly non-linear in their unknowns, which would make the solution of the system based on a maximum likelihood approach numerically intractable. However an approximate set of values $\bar{\mathbf{x}}$ can be given for all the parameters, thus the $j$-th AL observation equation can be linearized around these values
\begin{align}
	-\sin\phi_{j}^{\mathrm{c}}\,\delta\phi_{j}\simeq & \sum_{i=1}^{n^{\mathrm{S}}}
    	\left.\frac{\partial f_{\phi}\left(\mathbf{x}\right)}{\partial 
        	x_{i}^{\mathrm{S}}}\right|_{\bar{\mathbf{x}}}\delta 	
            x_{i}^{\mathrm{S}}+\sum_{i=1}^{n^{\mathrm{A}}}\left.
            \frac{\partial f_{\phi}\left(\mathbf{x}\right)}{\partial 
            x_{i}^{\mathrm{A}}}\right|_{\bar{\mathbf{x}}}\delta x_{i}^{\mathrm{A}}+
            \nonumber \\
 			& \sum_{i=1}^{n^{\mathrm{C}}}\left.\frac{\partial f_{\phi}
            \left(\mathbf{x}\right)}{\partial x_{i}^{\mathrm{C}}}
            \right|_{\bar{\mathbf{x}}}\delta x_{i}^{\mathrm{C}}+
            \sum_{i=1}^{n^{\mathrm{G}}}\left.\frac{\partial f_{\phi}\left(\mathbf{x}\right)}
            {\partial x_{i}^{\mathrm{G}}}\right|_{\bar{\mathbf{x}}}\delta x_{i}^{\mathrm{G}}
            \label{eq:linear_obs_eq}
\end{align}
where $\phi_{j}^{\mathrm{c}}=\arccos\left(f_{\phi}\left(\bar{\mathbf{x}}\right)\right)$, $\delta\phi_{j}=\phi_{j}^{\mathrm{o}}-\phi_{j}^{\mathrm{c}}$, $\delta x_{i}=x_{i}^{\mathrm{true}}-\bar{x}_{i}$, $\phi_{j}^{\mathrm{o}}$ is the abscissa of that observation measured by Gaia and $\mathbf{x}^{\mathrm{true}}$ is the set of unknown true values. A similar expansion can be written for the across scan measurements, and the problem can be thus reduced to the solution of a linear system of observation equations
\begin{equation}
	\mathbf{b}=A\,\delta\mathbf{x}\label{eq:eq_system_sphere}
\end{equation}
where $\mathbf{b}=\left\{b_j\right\} ^{\mathrm{T}}$, $j=1,\ldots,n_{\mathrm{obs}}$, is the vector of the known terms, $\delta\mathbf{x}$ is the unknown vector and $A$ is the $n_{\mathrm{obs}}\times n_{\mathrm{unk}}$ design matrix of the system whose coefficients are $a_{ji}=\left(\partial f/\partial x_{i}\right)\left(\bar{\mathbf{x}}\right)$. In the coefficients, obviously, $f()=f_{\phi}()$ or $f()=f_{\zeta}()$ if the observation is AL or AC respectively. The corresponding generic known term $b_j$ is equal to $-\sin\phi_{j}^{\mathrm{c}}\,\delta\phi_{j}$ or to $\cos\zeta_{j}^{\mathrm{c}}\,\delta\zeta_{j}$, and $n_{\mathrm{obs}}=n_{\mathrm{ALobs}}+n_{\mathrm{ACobs}}$, for the number of AL and AC observations $n_{\mathrm{ALobs}}$ and $n_{\mathrm{ACobs}}$ can in general be different.\footnote{The actual AL and AC observation equations are $\delta\phi_j=\left(-\sin\phi_{j}^{\mathrm{c}}\right)^{-1}\,\left[\sum_i a_{ji}\,\delta x_i\right]$ and $\delta\zeta_j=\left(\cos\zeta_{j}^{\mathrm{c}}\right)^{-1}\,\left[\sum_i a_{ji}\,\delta x_i\right]$ as this is a more convenient form for the post-fit analysis of the residuals.}

The resulting system of equations is intrinsically rank-deficient when astrometric and attitude parameters have to be solved at the same time. This well-known issue comes from the invariance of the solution for a rigid three-components rotation and three-components spin of the reference system, which in general requires the introduction of 6 constraint equations (see, e.g., discussions in \citet{1998A&A...332.1133D,2001A&A...373..336D,2016AJ....152...53B} about the actual need of such a constraint and their possible implementations.)

In order to take into account measurement errors, which are different for each observation, a $n_{\mathrm{obs}}\times n_{\mathrm{obs}}$ weight matrix\footnote{In this case, since the measurement errors can be considered independent at a first approximation, the weight matrix is the inverse square root of the covariance matrix of the known-term vector.}
\begin{equation}
	W=\mathrm{diag}\left(\frac{\epsilon_0}{\epsilon_j}\right)
\end{equation}
is introduced, where $\epsilon_j$ is the estimated standard deviation of the $j$-th observation and $\epsilon_0$ is a reference value, currently assumed to the estimated uncertainty at magnitude $G=21.5$. The system thus becomes
\begin{equation}
	W\mathbf{b}=(WA)\,\delta\mathbf{x}.\label{eq:eq_weighted_system}
\end{equation}

The solution in the least-squares sense of such system provides $\delta\mathbf{\hat{x}}$, namely the best-fit estimation of $\delta\mathbf{x}$, that is used to update the catalog values $\bar{\mathbf{x}}$ to the improved estimation of the true values $\tilde{\mathbf{x}}$, that is
\begin{equation}
	\mathbf{x}^{\mathrm{true}} \simeq \tilde{\mathbf{x}}=\bar{\mathbf{x}}+\delta\mathbf{\hat{x}}.
\end{equation}
The formal least-squares solution of the system of Eq.~(\ref{eq:eq_weighted_system}) is $\delta\mathbf{x}=\left[(WA)^{\mathrm{T}}(WA)\right]^{-1}(WA)^{\mathrm{T}}W\mathbf{b}$, where $\left[(WA)^{\mathrm{T}}(WA)\right]^{-1}$ is also the covariance matrix providing the estimation of the variances and covariances of the unknowns.

As mentioned in Section~\ref{sec:int_geod}, in this version of the pipeline the relativistic light deflection effect of Solar System objects different from the Sun is taken into account to improve on the model accuracy. However, it has to be stressed that this contribution is added only to the known term as a correction to $\phi_{j}^{\mathrm{c}}$ (and to $\zeta_{j}^{\mathrm{c}}$),\footnote{We remind that $\phi_{j}^{\mathrm{c}}$ and $\zeta_{j}^{\mathrm{c}}$ are computed in a purely Schwarzschild model.} and it is not considered in the derivatives. This is equivalent to a correction to the observations, which are thus forced to fit more closely the pure Schwarzschild model. The numerical consequence of this approach will be explained in Section~\ref{sec:DemonRun}.

\subsection{Solution of the linearized system of equations\label{sec:solsyst}}
It is known (see e.g.\ \citealt{2010A&A...516A..77B} and references therein) that the computational complexity of the inversion of an $N\times N$ matrix with direct methods is $\propto N^{3}$. Since the normal matrix $A^{\mathrm{T}}A$ is $n_{\mathrm{unk}}\times n_{\mathrm{unk}}$ with $n_{\mathrm{unk}}\sim6\cdot10^{8}$, we can immediately realize that in the Gaia case this problem can be addressed only resorting to iterative algorithms. A brute-force approach like this, in fact, would require about $10^{26}$ FLOPs a requirement which cannot be reduced to an acceptable level even taking into account the sparsity of $A$.

\begin{figure*}
	{\hfill{}\includegraphics[width=0.7\columnwidth]{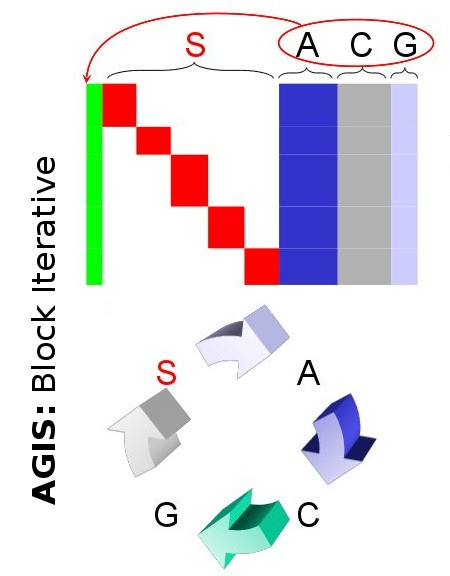}\hfill{}\includegraphics[width=0.7\columnwidth]{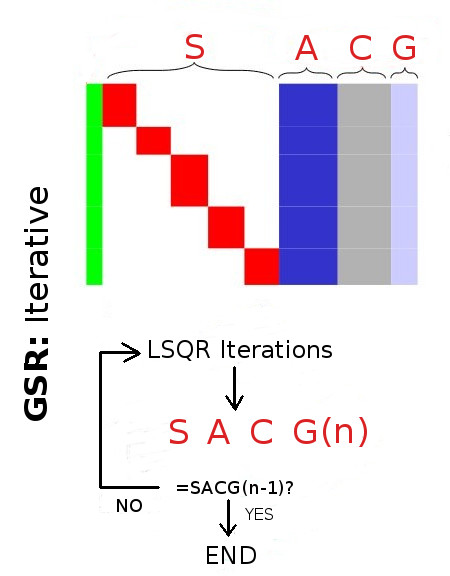}\hfill{}}
	\caption{\label{fig:block-vs-fully-iterative}Schematic representation of the block-iterative procedure used by AGIS (left) and of the fully iterative one used by GSR (right). See the text for explanation.}
\end{figure*}

AGIS uses a block-iterative technique to solve the equation system \citep{2012A&A...538A..78L} where, in practice, the solution is obtained by solving separately each block of unknowns ($\mathbf{x}^{\mathrm{S}}$, $\mathbf{x}^{\mathrm{A}}$, $\mathbf{x}^{\mathrm{C}}$ and $\mathbf{x}^{\mathrm{G}}$) and iterating the process until convergence (Fig.~\ref{fig:block-vs-fully-iterative}, left). This allows an implementation as an embarassingly parallel algorithm, which is strictly needed in the pure Java environment chosen in this case. On the other hand, the computation of standard uncertainties, and covariances, of the astrometric parameters requires the inversion of the entire normal matrix, which is needed to take properly into account the attitude-induced correlations among different sources; since this task is not feasible, one must rely on some approximate covariance model to estimate such statistical parameters \citep{2012A&A...543A..14H,2012A&A...543A..15H}.

Also lead by the initial consideration on the criticality of having an independent solution algorithm it was decided to use a fully iterative method (Fig.~\ref{fig:block-vs-fully-iterative}, right) to guarantee a fully general convergence mechanism of the complete system, and to allow a mixed Java/C-C++ coding of the pipeline in order to make possible the implementation of the needed parallel algorithm. Such implementation, in fact, is realized by means of a hybrid MPI/OpenMP parallel solver which runs at the CINECA supercomputing facilities \citep{BandieramonteBeccianiA.EtAl2012,Becciani2014}. GSR therefore implements a customized version of PC-LSQR \citep{Baur2008} a conjugate gradient-based algorithm, originally proposed by \citet{PaigeSaunders1982}. As for any iterative algorithm, this is equivalent to computing, at each iteration $\left(i\right)$, an approximate solution
\begin{equation}
	\delta\mathbf{x}^{\left(i\right)}=(A^{\mathrm{T}}A)^{-1}A^{\mathrm{T}}\mathbf{b}^{\left(i-1\right)}
\end{equation}
and then evaluating the vector of the residuals
\begin{equation}
	\mathbf{r}^{\left(i\right)}=\mathbf{b}-A\,\delta\mathbf{x}^{\left(i\right)}
\end{equation}
which has to be minimized in the least-squares sense, according to suitable convergence conditions defined by the algorithm itself. Among the possible stopping conditions we have:
\begin{itemize}
	\item the vector of the residuals has a 2-norm lower than a threshold value; the LSQR algorithm generates a series of residual vectors whose norm decreases monotonically; in the case of a compatible system the series goes to zero, while for non-compatible systems it converges to a positive finite limit;
    \item the 2-norm of $A^\mathrm{T}\mathbf{r}^{\left(i\right)}$ (namely the norm of the residuals of the normal system $A^\mathrm{T}\mathbf{b}=A^\mathrm{T}A\delta\mathbf{x}$) is lower than a threshold value; this is the condition used to guarantee the convergence of non-compatible systems to a solution in the least-squares sense; it is worth stressing here that the equation system to be solved for the global astrometric sphere reconstruction is non-compatible,\footnote{That is, it has no exact solutions, namely solutions for which $\mathbf{r}^{\left(i\right)}=0\ \forall\ i$.} and indeed its solution is obtained in the least-squares sense, therefore this is the stopping condition that has to be reached in our case;
    \item the iterative estimation of the condition number of the matrix exceeds a given upper threshold;
	\item a fixed maximum number of iterations is reached.
\end{itemize}
The choice of the LSQR algorithm is also motivated by the possibility of further enhancing its standard implementation. Indeed, in its original definition it provides an estimation only of the diagonal elements of the covariance matrix $\left(A^{\mathrm{T}}A\right)^{-1}$, namely the variances $\sigma_{\delta\mathbf{x}}$, but upgrades of this algorithm (see, e.g. \citealt{guo_diplomarbeit_2008,Kostina_and_Kostyukova-2012}) would allow to estimate also any selected group of its off-diagonal elements, namely the covariances.

The GSR version of the algorithm uses a preconditioning technique, which basically consists in a renormalization of the columns of $A$, made to improve the convergence speed of the system. This version, moreover, is tailored on the sphere solution problem in particular for what concerns the parallelization algorithm and the memory occupancy. The latter has been optimized with respect to the classic Yale Sparse Matrix Format \citep{BulucFinemanFrigoEtAl2009} exploiting the almost constant number of non-zero coefficients to eliminate one of the $n_{\mathrm{obs}}$ pointer vector. Regarding the former, instead, the design matrix is built by sorting the observation by source number. In this way it is possible to distribute the stars in almost independent subsets on each Processing Element (PE), in fact the matrix-vector product, which is the core of the LSQR algorithm, can be computed by distributing on each PE an equal number $n_{\mathrm{obs}}/n_{\mathrm{PE}}$ of the rows of $A$ and of the vector \textbf{$\mathbf{b}$, }and a number $5n^{*}/n_{\mathrm{PE}}$ of astrometric unknowns. The other unknowns are then duplicated on all the PEs making the product almost communication-free \citep{BandieramonteBeccianiA.EtAl2012}.

\subsection{Differential attitude and AC observations}
Let us say that one has to compute the attitude coefficients of a generic linearized observation equation. From the above considerations we can formally write
\begin{equation}
	\mathbf{x}^{\mathrm{A}}=\left\{ \sigma_{1}\left(c_{0}^{
    	\left(1\right)},\ldots,c_{N-1}^{\left(1\right)}\right),
        \sigma_{2}\left(c_{0}^{\left(2\right)},\ldots,
        c_{N-1}^{\left(2\right)}\right),\sigma_{3}
        \left(c_{0}^{\left(3\right)},\ldots,c_{N-1}^{\left(3\right)}\right)\right\},
\end{equation}
so that in general the catalog attitude is a set of initial values of the B-Spline expansion coefficients $\left\{ \bar{c}_{i}^{\left(j\right)}\right\}$. Eq.~(\ref{eq:linear_obs_eq}) can thus be written as
\begin{align}
	-\sin\phi_{\mathrm{calc}}\,\delta\phi\simeq & \sum_{i=1}^{n^{\mathrm{S}}}
    	\left.\frac{\partial f_{\phi}\left(\mathbf{x}\right)}{\partial 
        	x_{i}^{\mathrm{S}}}\right|_{\mathbf{x}_{0}^{\mathrm{S}}}\delta 
            x_{i}^{\mathrm{S}}+\sum_{j=1}^{3}\left.\frac{\partial f_{\phi}\left(\mathbf{x}\right)}
            {\partial\sigma_{j}}\right|_{\bar{\sigma}_{j}}\delta\sigma_{j}+\nonumber \\
 		& \sum_{i=1}^{n^{\mathrm{C}}}\left.\frac{\partial f_{\phi}\left(\mathbf{x}\right)}{\partial 
        	x_{i}^{\mathrm{C}}}\right|_{\bar{\mathbf{x}}}\delta 
            x_{i}^{\mathrm{C}}+\sum_{i=1}^{n^{\mathrm{G}}}
        \left.\frac{\partial f_{\phi}\left(\mathbf{x}\right)}{\partial 
        x_{i}^{\mathrm{G}}}\right|_{\bar{\mathbf{x}}}\delta x_{i}^{\mathrm{G}}\nonumber \\
	= & \sum_{i=1}^{n^{\mathrm{S}}}\left.\frac{\partial f_{\phi}
    	\left(\mathbf{x}\right)}{\partial x_{i}^{\mathrm{S}}}\right|_{\mathbf{x}_{0}^{\mathrm{S}}}\delta 
        x_{i}^{\mathrm{S}}+\sum_{j=1}^{3}\sum_{i=0}^{N-1}
        \left.\frac{\partial f_{\phi}\left(\mathbf{x}\right)}{\partial\sigma_{j}}
        \frac{\partial\sigma_{j}}{\partial c_{i}^{\left(j\right)}}
        \right|_{\bar{c}_{i}^{\left(j\right)}}\delta c_{i}^{\left(j\right)}\nonumber \\
	 & \sum_{i=1}^{n^{\mathrm{C}}}\left.\frac{\partial f_{\phi}\left(\mathbf{x}\right)}
     	{\partial x_{i}^{\mathrm{C}}}\right|_{\bar{\mathbf{x}}}\delta 
        	x_{i}^{\mathrm{C}}+\sum_{i=1}^{n^{\mathrm{G}}}
        	\left.\frac{\partial f_{\phi}\left(\mathbf{x}\right)}{\partial 
            x_{i}^{\mathrm{G}}}\right|_{\bar{\mathbf{x}}}\delta x_{i}^{\mathrm{G}}\nonumber \\
	= & \sum_{i=1}^{n^{\mathrm{S}}}\left.\frac{\partial f_{\phi}\left(\mathbf{x}\right)}
    	{\partial x_{i}^{\mathrm{S}}}\right|_{\mathbf{x}_{0}^{\mathrm{S}}}\delta 
        	x_{i}^{\mathrm{S}}+\sum_{j=1}^{3}\sum_{i=0}^{N-1}
            \left.\frac{\partial f_{\phi}\left(\mathbf{x}\right)}
            {\partial\sigma_{j}}\right|_{\bar{\sigma}_{j}}B_{i}\left(t\right)
            \delta c_{i}^{\left(j\right)}\nonumber \\
 	& \sum_{i=1}^{n^{\mathrm{C}}}\left.\frac{\partial f_{\phi}
    	\left(\mathbf{x}\right)}{\partial x_{i}^{\mathrm{C}}}\right|_{\bar{\mathbf{x}}}\delta 
        	x_{i}^{\mathrm{C}}+\sum_{i=1}^{n^{\mathrm{G}}}\left.\frac{\partial 
            	f_{\phi}\left(\mathbf{x}\right)}{\partial x_{i}^{\mathrm{G}}}\right|_{\bar{\mathbf{x}}}\delta 
            x_{i}^{\mathrm{G}}\label{eq:att_coeff_barsigma}
\end{align}
in which we have exploited the fact that
\begin{equation}
	\mathrm{d}\sigma_{j}\left(t\right)=\sum_{i=0}^{N-1}
    	\frac{\partial\sigma_{j}}{\partial c_{i}^{\left(j\right)}}\mathrm{d}c_{i}^{\left(j\right)}
\end{equation}
and that, from Eq.~(\ref{eq:MRP_bspline}),
\begin{equation}
	\frac{\partial\sigma_{j}}{\partial c_{i}^{\left(j\right)}}=B_{i}\left(t\right).
\end{equation}

In this way the updated attitude at a generic time $t$ is
\begin{equation}
	\sigma_{j}(t)=\sum_{i=0}^{N-1}
	\left(\bar{c}_{i}^{\left(j\right)}+\mathrm{\delta}c_{i}^{(j)}\right)
    	B_{i}\left(t\right).\label{eq:complete_attitude_update}
\end{equation}

It should be observed, however, that in Eq.~(\ref{eq:att_coeff_barsigma}) $\partial f_{\phi}\left(\mathbf{x}\right)/\partial\sigma_{j}$ does not depend on the representation of $\mathrm{d}\sigma_{j}$: it just needs a (catalog) value for the attitude parameter $\bar{\sigma}_{j}\equiv\bar{\sigma}_{j}\left(t\right)$ at the observation time $t$. We can therefore decide to compute this $\sigma_{j}$ using an expansion which, in principle, may have nothing to do with the one used to represent $\mathrm{d}\sigma_{j}$ since our only necessity is to have a way to evaluate $\bar{\sigma}_{j}\left(t\right)$. In formulae, we could write at any time $\sigma_{j}(t)=\bar{\sigma}_{j}\left(t\right)+\mathrm{d}\sigma_{j}\left(t\right)$ where
\begin{align}
	\bar{\sigma}_{j}\left(t\right) & 
    	=\sum_{m=0}^{N_{\mathrm{C}}-1}c_{m,\mathrm{C}}^{\left(j\right)}B_{m}\left(t\right)\\
	\mathrm{d}\sigma_{j}\left(t\right) & 
    	=\sum_{i=0}^{N_{\mathrm{U}}-1}c_{i,\mathrm{U}}^{\left(j\right)}B_{i}\left(t\right)
\end{align}
are the catalog attitude and its update, exactly as in our original scenario, but expanded over two different supports $\left\{ \tau_{m,\mathrm{C}}\right\} $, $m=0,\ldots,N_{\mathrm{C}}$ and $\left\{ \tau_{i,\mathrm{U}}\right\} $, $i=0,\ldots,N_{\mathrm{U}}$. Moreover, since the catalog attitude has to represent the starting point for the unknowns, we should have initially $\mathrm{d}\sigma_{j}\left(t\right)=0$, that is $c_{i,\mathrm{U}}^{\left(j\right)}=0$, so that in the end $c_{i,\mathrm{U}}^{\left(j\right)}=\delta c_{i,\mathrm{U}}^{\left(j\right)}$, and therefore the updated attitude is computed at any time $t$ simply by summing the catalog and the differential attitude $\mathrm{d}\sigma$ at the same time:
\begin{eqnarray}
	\sigma_{j}(t) & = & \bar{\sigma}_{j}\left(t\right)+\mathrm{d}\sigma_{j}\left(t\right)\\
		& = & \sum_{m=0}^{N_{\mathrm{C}}-1}c_{m,\mathrm{C}}^{\left(j\right)}
  				B_{m}\left(t\right)+\sum_{i=0}^{N_{\mathrm{U}}-1}
                \delta c_{i,\mathrm{U}}^{\left(j\right)}B_{i}\left(t\right).
\end{eqnarray}
This facilitates the choice of the knot support in the computation of the coefficients and makes the calculation of the latter much faster. The placement of the knots, in fact, has to guarantee that each attitude parameter can be solved, a condition that can be met basically by allowing a sufficient number of observations between each pair of consecutive knots. There are, however, several reasons which could require to drop some observations after the computation of the matrix coefficients, thus inducing a rearrangement of the knot sequence; in the non-differential approach, such rearrangement would entail a recalculation of all the coefficients, while in the differential approach the latter must be computed only once, with the exception of the $B_{i}\left(t\right)$ polynomials for the differential knot sequence.

One last point of the Gaia attitude reconstruction has to be stressed. Given the three coefficients $C_{k}^{\left(j\right)}$, $j=1,2,3$ relative to the $k$-th unknown of each Rodrigues parameter $\sigma_{j}$,\footnote{The considerations we are writing for the MRP are valid for any representation which is expanded in B-Splines series.} their ratios $R_{k}^{\left(i,j\right)}=C_{k}^{\left(i\right)}/C_{k}^{\left(j\right)}$ do not depend from the B-spline base function $B_{k}\left(t\right)$. As a consequence, whatever the time dependence of two coefficients, one can always find a small time separation $\delta t=t_{2}-t_{1}$ between two observations at $t_{1}$ and $t_{2}$ respectively which is short enough to keep $R_{k}^{\left(i,j\right)}$ almost constant between $t_{1}$ and $t_{2}$. As pointed out above, this ``constancy'' in principle does not depend on the explicit formulation of the coefficients, since it is sufficient to have time-dependent expressions repeating themselves along all the observations, however it is easy to understand that the slower the time variation, the longer this time interval can be kept. Therefore the independence of the ratios from $B_{k}\left(t\right)$ contributes to increase the amount of time over which the $R_{k}^{\left(i,j\right)}$ remain almost the same among the different observations or, equivalently, over which the proportionality 
\(
\mathbf{C}_{k}^{\left(1\right)} \propto
\mathbf{C}_{k}^{\left(2\right)} \propto
\mathbf{C}_{k}^{\left(3\right)}
\)
holds approximately for each index $k$.

Since each of these values is represented into a different column of the system of equations, this quasi-proportionality produces an extremely ill-conditioned design matrix. Such an ill-conditioning problem is solved by introducing the AC observations. This problem will be discussed in more detail in a forthcoming publication (\citet{2019Models...InPrep..V}, in preparation).

\begin{figure}
	\includegraphics[width=\hsize]{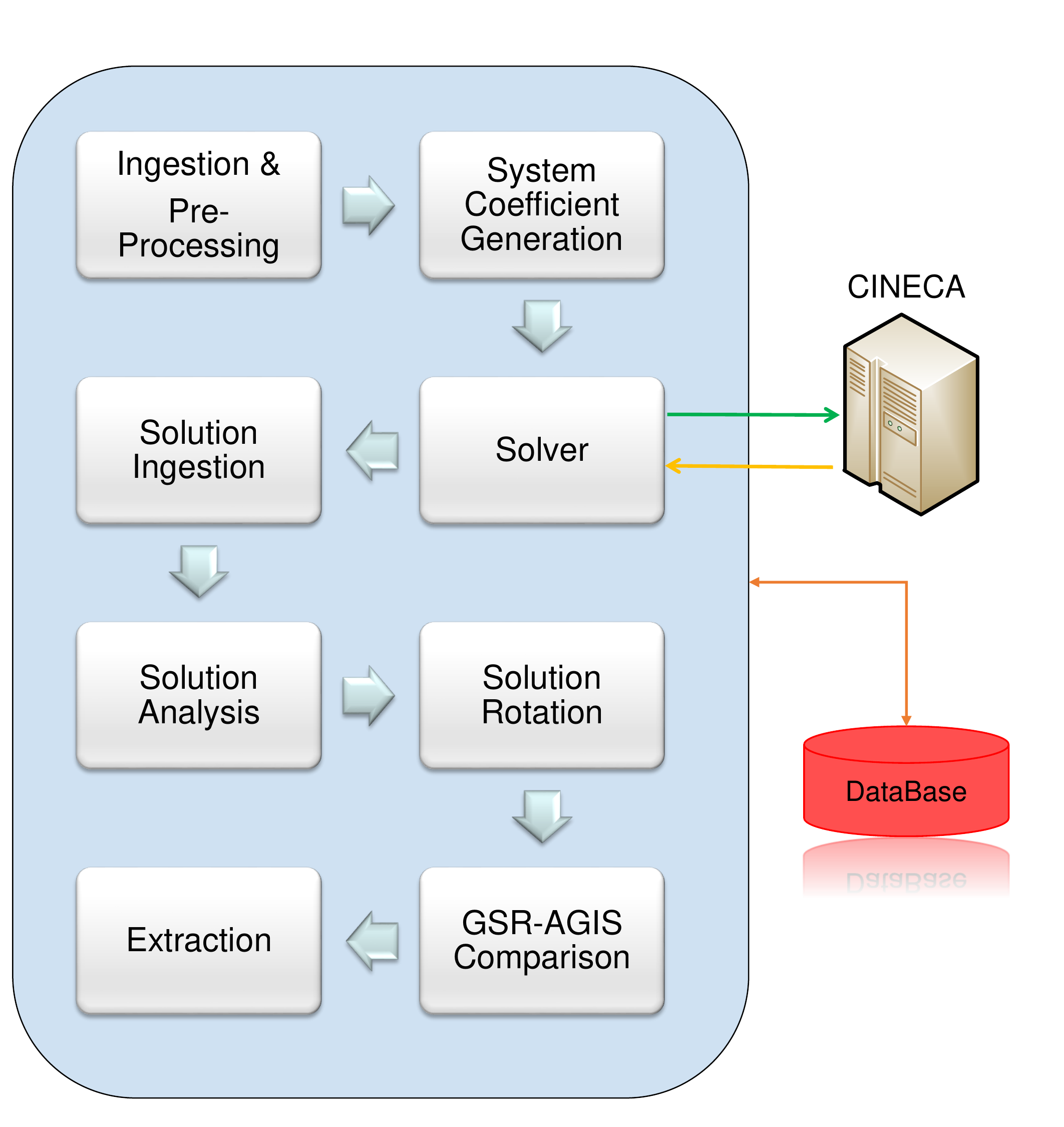}
	\caption{\label{fig:GSR_pipeline}Schematic representation of the AVU/GSR pipeline.}
\end{figure}

\section{The AVU/GSR pipeline}
\subsection{The overall pipeline schema}
As mentioned in Section~\ref{sec:Introduction} the algorithms described so far have been implemented in the GSR pipeline, which operates in the context of the Astrometric Verification Unit of the DPAC. A detailed description of such pipeline is out of the scope of this paper, and more details can be found in \citet{VecchiatoetalSPIE2012} and in a forthcoming publication (\citealt{2019GSRPipeline...InPrep..M}, in preparation). Here we will thus give a synthetic summary of its main characteristics.

The GSR pipeline is composed of several scientific modules, called in sequence by the infrastructure software, which provides the overall workflow and the DB interaction functionalities (Figure~\ref{fig:GSR_pipeline}). In essence, GSR gets the AGIS solution and its corresponding input data. The latter are used to produce an independent sphere solution which is then compared to that of AGIS. The results of the comparison, as outcomes of statistical tests and graphics, are sent to the Gaia Main DataBase (MDB). Alerts are also foreseen if the comparison bears evidence of statistically significant differences between the two solutions.

All the scientific modules can be grouped in three main types:
\begin{itemize}
	\item The ``Pre-solver'' modules, which implements the computation of the system coefficients and known terms according to the guidelines sketched in the Sections~\ref{sec:Modeling-the-observations} to \ref{sec:Reconstructing}.
	\item The ``Solver'' module, which computes the system solutions as described in Section~\ref{sec:Reconstructing}
	\item The ``Post-solver'' modules, performs all the successive operations from the catalog update to the comparison between the AGIS and GSR solutions summarized in Section~\ref{subsec:Comparison}.
\end{itemize}
As anticipated, all the modules are written in Java and run at the DPCT-ALTEC, with the exception of the Solver module which is written in C/C++ and runs at the DPCT-CINECA. The latter simply receives the coefficients and known terms for the system from ALTEC and sends back there the system solution. The pipeline concludes with the production of a report which contains all the relevant information needed to give a first evaluation of the results.

\subsection{Post-Solver GSR pipeline modules\label{subsec:Comparison}}
As mentioned above, GSR has to compare its results with the AGIS solution and to provide an evaluation of the differences between the two spheres. In particular, it is foreseen that possible differences larger than the expected accuracy of Gaia are investigated to assess the scientific reliability of the result.

As pointed out in Section~\ref{sec:Reconstructing} the sphere reconstruction problem is intrinsically rank-deficient, and as a minimum it can be solved except for a six-parameters transformation representing a rigid rotation and a spin difference of the reference system. In the equation system solved by GSR, these can be directly included as additional constraint equations, equivalent to the choice of a specific (and arbitrary) reference system which, in general, is different from that of the AGIS solution. It is therefore necessary to de-rotate one solution, that is to bring both catalogs into a common reference system, before attempting a comparison between them. Moreover, both the de-rotation and comparison procedures are done on catalogs referring to the purely spatial hypersurface of the same observer, therefore there is no need to resort to General Relativity here. The algorithms are least-squares reconstructions of purely Euclidean transformations and statistical analysis of coordinate differences.

GSR can use standard statistical algorithms, like $\chi^{2}$ tests and Kolmogorov-Smirnov, to analyze the differences between the AGIS and GSR global astrometric sphere solutions. In addition to these, a powerful method in this context is provided by the use of a decomposition in Scalar and Vector Spherical Harmonics (VSH, see e.g.\ \citealt{Hill-1954,Arfken2012}) as base functions to model the vector field of GSR/AGIS residuals as a series expansion, up to a suitable degree, whose coefficients can be either estimated by a standard least-squares fit or computed analytically, whenever appropriate. Once the significance of the various coefficients associated to each VSH degree is positively tested, the latter quantify the presence of a systematic error in the residuals at scale lengths of the order of $180/l$~deg, where $l$ is the corresponding VSH degree, therefore their nature must be addressed.

This last algorithm is the most computationally-intensive, so its implementation is the only one besides the system solution that can require a non-embarrassingly parallel coding (to be executed at the DPCT-CINECA) in case a decomposition of order $l_{\mathrm{max}}\gtrsim 100$ is required. The details of the algorithms can be found in \citet{2011LL...BB00101}, \citet{2011LL...BB00201} and in a forthcoming paper (\citealt{2019Comparison...InPrep..B}, in preparation), which also contains a characterization of the VSH accuracy for the Gaia case.

\section{Demonstration runs on simulated data\label{sec:DemonRun}}
\subsection{General considerations}
Similarly to what was done for AGIS \citep{2012A&A...538A..78L}, the accuracy of the AVU/GSR pipeline has to be verified before this software system enters operations. The general agreement was that GSR had to perform a ``Demonstration Run test,'' showing its ability to reproduce the AGIS results at a comparable accuracy level on simulated data similar to those used in the cited paper. For that purpose, the GSR demonstration run was split into two different tests:
\begin{description}
	\item [{Test~1:}] solve the GSR astrometric sphere with perturbed Source (S), Attitude (A) and Calibration (C) parameters and noise-free observations. The aim is to give a precise assessment of the numerical accuracy of the AVU/GSR pipeline and of the GSR2 astrometric model.
    \item [{Test~2:}] solve the GSR astrometric sphere with perturbed S, A and C parameters and noisy observations, similarly to what was done for AGIS (see the cited paper above). The goal is reproducing the AGIS results at the appropriate accuracy.
\end{description}
Following the methodology adopted in the AGIS paper, the perturbations of the source and attitude parameters are Gaussian, while the system starts from true values for the instrument (calibration parameters). Moreover, a long-period modulation of the BA is injected in the known terms (observations) and reconstructed by means of the $L_{0}$ large-scale AL Instrument parameters ($\Delta\eta$) at the sub-as level. The instrument model is used only in this scope.

As shown in Sect.~\ref{sec:Reconstructing}, the global astrometric sphere reconstruction, in the case of Gaia, requires the solution in the least-squares sense of a large, sparse and overdetermined system of linearized equations. In GSR the resulting linear equation system is solved in full by utilizing a parallelized implementation of the LSQR algorithm. Provided that the starting values are sufficiently close to the true ones the system converges to its best possible least-squares solution, a statement which has to be intended in comparison with the block-iterative algorithm adopted by AGIS.

This simple framework is complicated by the discrepancy between the astrometric accuracy of the two relativistic models implemented in AGIS and the current GSR. AGIS implements the GREM astrometric model, which is accurate to the $\left(v/c\right)^{3}$ order needed to match the Gaia mission final accuracy, and the simulated data are generated with the same astrometric model. The current GSR, instead, uses a model of the RAMOD family whose $\left(v/c\right)^{2}$ PPN-Schwarzschild metric, as explained in Section~\ref{sec:int_geod}, is improved in accuracy by an approximate estimation of the contributions of the planets of the solar system and of the Moon, which are subtracted from the known terms.

This technique, however, cannot model at the required level of accuracy the observation equation of this intrinsically non-Schwarzschild problem. The accuracy of the GSR2 astrometric model, therefore, does not match that of the simulated data everywhere on the sky, and leaves unmodeled some effects at the order of the first derivative of the observable, that enter directly in the sphere reconstruction problem. Furthermore, and more importantly, it does not cope with the final accuracy expected for Gaia, especially at the bright end of the magnitude range. This approximate approach could be further improved, but at the price of a considerable mathematical complication, an effort that in any case would not be able to ensure a perfect match in terms of model accuracy.

On the other hand, it is still possible to reach (almost) the same results of the AGIS demonstration run by minimizing the impact of these modeling issues by, firstly, dropping observations too close to the planets and the Moon, and secondly, starting as close as possible to the true values. The second requirement can be met with an ad-hoc procedure, adopted for both tests constituting the demonstration run, that consists of three steps:
\begin{enumerate}
	\item The first is a regular run of the GSR pipeline starting from input values comparable to those of AGIS.
	\item The second starts from the same input values, but uses the BA reconstruction of the first step to remove the modulation from the known terms.
	\item The third and final step is another run of the GSR pipeline in which the input attitude is the same as that in step 1, while the input values for the astrometric parameters are those of the solution of step 2 (the step 2 residual BA modulation is also subtracted from the known term as before).
\end{enumerate}
We stress that, as long as we consider the simulated data a faithful reproduction of reality, the adopted procedure is not motivated just by the accuracy mismatch between the two astrometric models, but also by that between the accuracy of the RAMOD model and that of the Gaia measurements. Moreover, the third step takes into account the residual model-induced inaccuracy of the first order derivatives of the observable and for this reason we dub it ``external iteration,'' or EI run.

Indeed, it is worth noticing that this run might look like the so-called iteration for non-linearity, but it has actually another meaning. Indeed, the former is a standard procedure used when the input values are so approximate that the second order effects neglected in the linearized problems are still significantly large with respect to the measurement accuracy. Our EI, instead, is a numerical procedure that takes into account a first-order modelling accuracy issue. In short, it comes from the fact, already mentioned in Sect.~\ref{sec:LinEqSyst}, that it was chosen to model the effects of the planets only for the known terms, not including them in the coefficients of the linearized observation equation. This was done for practical reasons, and a detailed explanation, which would be too long and out of scope here, will be provided in a forthcoming paper \citep{2019Models...InPrep..V}. In this respect, when the current model will be replaced by a full-accuracy astrometric model \citep{2017PhRvD..96j4030C,2017A&A...608A..83B} this ad-hoc procedure will become unnecessary. GSR, in fact, will have the same sensibility of AGIS to modelling errors and will match the Gaia measurement accuracy, so in principle it will be able to reach the required accuracy just after the first step.

The  dataset utilized in these tests provides the true Gaia AL and AC measurements, which are computed from the coordinates of the simulated objects, the ephemerides of the planets and of the satellite, the NSL, and from the unperturbed instrument parameters using the implementation of the GREM model available when this dataset was produced. In comparing the results of our tests with those of AGIS, it should be considered that the former used a dataset containing 908,979 primary sources in the magnitude range $5.79<G<20.00$, while AGIS, in its demonstration run, could use another version of the same dataset, not available anymore, that had more than 2 million primaries \citep{2012A&A...538A..78L}.

The perturbed values needed for the tests are generated with two different procedures for the sources and the attitude. The source parameters are perturbed with a Gaussian noise with zero average and a standard deviation of 20~mas for positions and annual proper motions. Moreover, negative parallaxes are not admitted, and when a random extraction produces a negative value, it is set to $\varpi=10^{-6}\:\mathrm{mas}\simeq4.8\times10^{-15}\:\mathrm{rad}$.\footnote{This is just a practical recipe to obtain sensible values for known terms and coefficients in the GSR formulae. After a solution is found, both positive and negative values are admitted for the updated parallaxes.} The starting attitude values, instead, are obtained by perturbing the coefficients of the B-splines fit of the NSL (true attitude), with a Gaussian noise that produces a difference of about 10~mas between the true and perturbed orientations of the attitude axes. The initial separation between two successive nodes of the knot sequence is set to 240~s.

In test 2, the one with noisy observations, the perturbed observations are obtained by adding a Gaussian measurement noise to the true measured values. All the information needed to compute the latter are contained in the simulated dataset. The Gaussian perturbation, instead, is computed by generating a random extraction for each observation from a Gaussian distribution centered in zero and having a standard deviation corresponding to the observational error. This depends on the magnitude of the observed star, and it is obtained using the DPAC routines that implement the nominal AL/AC single-measurement error.

A star is declared solvable (for all of the 5 astrometric parameters) if it has at least 180 AF along-scan observations, and, at the same time, the difference between the observation times of its first and last observation is at least 1.5 years.

An important part of the coefficients module is the Attitude Definition Chain (ADC). The task of this piece of software is the definition of the knot sequence of the attitude, whose basic nominal separation can be adjusted to fit some constraints. Basically, there must be at least 20 observations between two adjacent knots, and the interval can be stretched up to four times the initial knot separation, namely 960~s. If after this time the minimum number of observations has not been reached yet, the B-spline sequence is segmented, thus a discontinuity in the attitude reconstruction is introduced.

The convergence condition of the LSQR algorithm are set to the most stringent requirement, namely to the machine precision accuracy. This condition is overridden if the estimated condition number of the system exceeds $10^{13}$, the number of iterations is larger than 50,000, or if the solver runs for more than 120 hours. The preconditioning of the equation system is activated (to speed up convergence), and the six constraint equations needed to fix the intrinsic rank deficiency of the system are computed from the catalog values of ``one-and-a-half stars.'' Namely, we first select all stellar pairs according to the following geometric criteria:
\begin{align}
	\left|\delta_{1,2}\right| & <5\:\mathrm{deg}\\
	90-5\:\mathrm{deg}<\left|\alpha_{2}-\alpha_{1}\right| & <90+5\:\mathrm{deg}.
\end{align}
Final choice is done by choosing the brightest two among the above pairs, and the constraint equations are built by fixing $\alpha$, $\delta$, $\mu_{\alpha}$ and $\mu_{\delta}$ of the brightest and $\delta$ and $\mu_{\delta}$ of the second brightest.\footnote{It is worth stressing that, in principle, the LSQR algorithm does not necessarily need any constraint equations to converge to one of the least-squares solution of the system. In this case, however, the convergence would be much slower.}

\subsection{Test 1: accuracy assessment with noise-free observations}
\begin{figure*}
	\resizebox{\hsize}{!}
	{\includegraphics{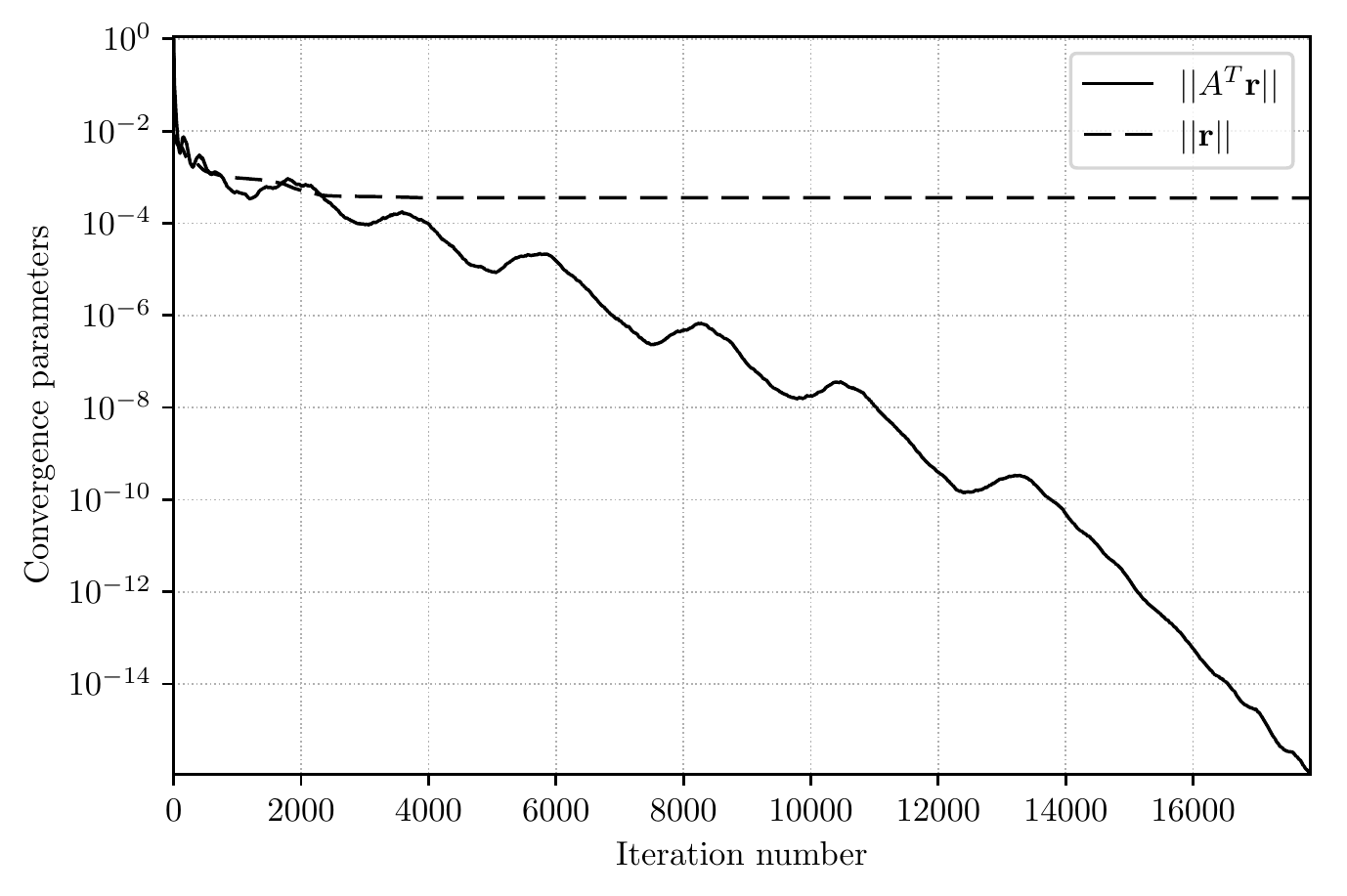}\hfill{}\includegraphics{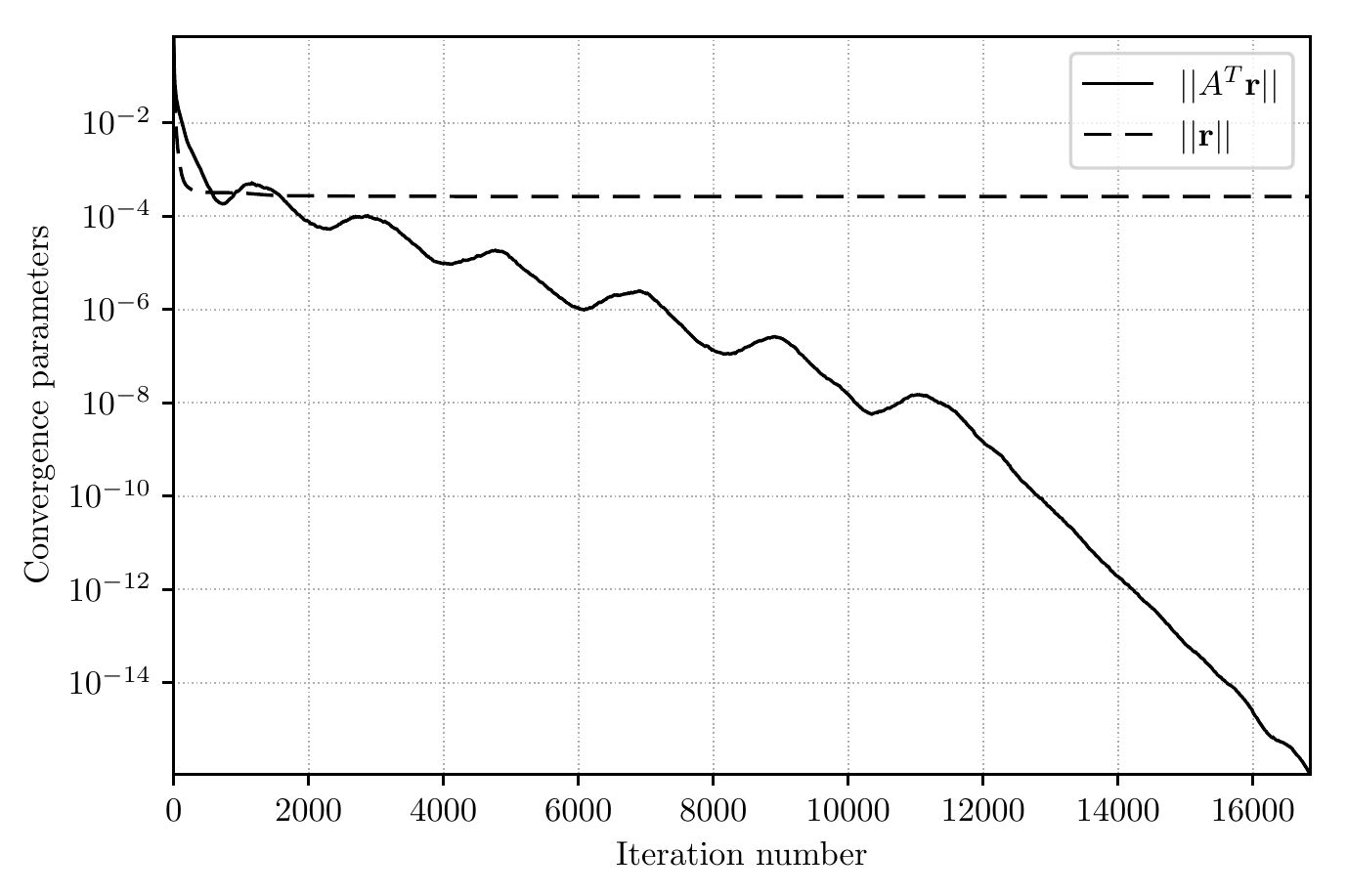}}
	\caption{\label{fig:Convergence-plots-1}Convergence plots of the first and third step. The LSQR algorithm implementation monitors the convergence status of the solution by computing two parameters, the 2-norm of the residuals vector ($||\mathbf{r}||\equiv||A\mathbf{x}-\mathbf{b}||$), and the 2-norm $||A^T\mathbf{r}||$ (dashed and solid lines respectively). As mentioned in Section~\ref{sec:solsyst}, convergence to the unique least-squares solution is confirmed by the fact that the LSQR estimation of $||A^T\mathbf{r}||$ is zero within the machine-precision accuracy.}
\end{figure*}

The GSR pipeline filters out 460 primary sources that cannot be solved because they do not fit the minimum requirements explained in the previous section. Therefore, this leaves a system with $908\,519$ primaries and $660\,599$ degrees of freedom for the knot sequence. The attitude is represented in terms of MRP which implies that we have just 3 independent unknowns per knot, instead of the 4 constrained ones of the quaternion representation. In addition, the large scale instrument parameters that are estimated in the solution are associated to a total of 63 CCDs, each varying with a time scale of one month.

In this representation, therefore, the number of astrometric unknowns is $4\,542\,595$, that of the attitude is $1\,981\,797$, and finally, the number of AL and AC (large scale) instrument parameters is $45\,360$. The linearized system of equations is thus described by a design matrix of $6\,569\,752$ columns (unknowns) and $1\,330\,628\,523$ rows (observations). In each of the three steps of this test, the solver converged to the machine-precision least-squares solution in something less than $20\,000$ iterations, with a condition number of $\sim10^{6}$ (Figure~\ref{fig:Convergence-plots-1}).

\begin{table*}
\caption{Astrometric results (estimated minus true) for test 1. Units are in $\mu\mathrm{as}$ for parallaxes and positions and $\mu\mathrm{as}/\mathrm{yr}$ for proper motions. Right ascension and the corresponding proper motion are provided as $\alpha^*=\alpha\cos\delta$ and $\mu_{\alpha^*}=\mu_\alpha\cos\delta$.}
\label{tab:Astrometric-results-gsr_test-1}      
\centering          
\begin{tabular}{c c c c c c c c c c c c }  
\hline\hline
  \multirow{2}{*}{Magnitude range} & \multirow{2}{*}{Step} & \multicolumn{5}{c}{Median} & \multicolumn{5}{c}{RSE} \\
  & & $\varpi$ & $\alpha^{*}$ & $\delta$ & $\mu_{\alpha^{*}}$ & $\mu_{\delta}$ & $\varpi$ & $\alpha^{*}$ & $\delta$ & $\mu_{\alpha^{*}}$ & $\mu_{\delta}$ \\
\hline
  \multirow{2}{*}{$\phantom{13\leq}G<13$}  & 1 & -0.78 & 4.09 & 1.40 & 4.61 & -1.00 & 2.79 & 10.03 & 7.00 & 12.93 & 5.26 \\
  	                                       & 3 &  0.22 & 0.01 & 0.07 & 0.00 &  0.02 & 0.17 &  0.38 & 0.49 & 0.10 & 0.10 \\
  \cline{2-12}
  \multirow{2}{*}{$13\leq G<15$}           & 1 & -0.76 & 3.68 & 1.41 & 4.61 & -1.16 & 3.08 &  9.63 & 6.66 & 11.79 & 4.62 \\
  	                                       & 3 &  0.22 & 0.01 & 0.05 & 0.01 &  0.03 & 0.17 &  0.21 & 0.29 &  0.09 & 0.09 \\
  \cline{2-12}
  \multirow{2}{*}{$15\leq G<16$}           & 1 & -0.67 & 3.42 & 1.26 & 4.51 & -1.32 & 3.32 &  9.76 & 6.55 & 12.09 & 4.27 \\
  	                                       & 3 &  0.24 & 0.01 & 0.05 & 0.01 &  0.03 & 0.17 &  0.18 & 0.22 &  0.08 & 0.09 \\
  \cline{2-12}
  \multirow{2}{*}{$16\leq G<17$}           & 1 & -0.60 & 3.38 & 1.16 & 4.59 & -1.36 & 3.37 &  9.95 & 6.33 & 11.76 & 4.14 \\
  	                                       & 3 &  0.25 & 0.01 & 0.04 & 0.01 &  0.03 & 0.16 &  0.17 & 0.19 &  0.08 & 0.09 \\
  \cline{2-12}
  \multirow{2}{*}{$17\leq G<18$}           & 1 & -0.53 & 3.25 & 1.09 & 4.57 & -1.40 & 3.37 & 10.12 & 6.16 & 11.59 & 4.04 \\
  	                                       & 3 &  0.26 & 0.00 & 0.04 & 0.01 &  0.03 & 0.16 &  0.16 & 0.16 &  0.08 & 0.09 \\
  \cline{2-12}
  \multirow{2}{*}{$18\leq G<19$}           & 1 & -0.42 & 3.10 & 0.90 & 4.56 & -1.50 & 3.49 & 10.22 & 6.02 & 11.46 & 3.84 \\
  	                                       & 3 &  0.27 & 0.00 & 0.04 & 0.01 &  0.04 & 0.16 &  0.15 & 0.15 &  0.08 & 0.08 \\
  \cline{2-12}
  \multirow{2}{*}{$19\leq G\phantom{<15}$} & 1 & -0.18 & 3.01 & 0.52 & 4.65 & -1.63 & 3.59 & 10.40 & 5.68 & 11.62 & 3.59 \\
  	                                       & 3 &  0.29 & 0.00 & 0.04 & 0.01 &  0.04 & 0.16 &  0.15 & 0.14 &  0.08 & 0.08 \\
\hline
\end{tabular}
\end{table*}

\begin{figure*}
    \centering
	\resizebox{0.5\hsize}{!}
    {\hfill{}\includegraphics{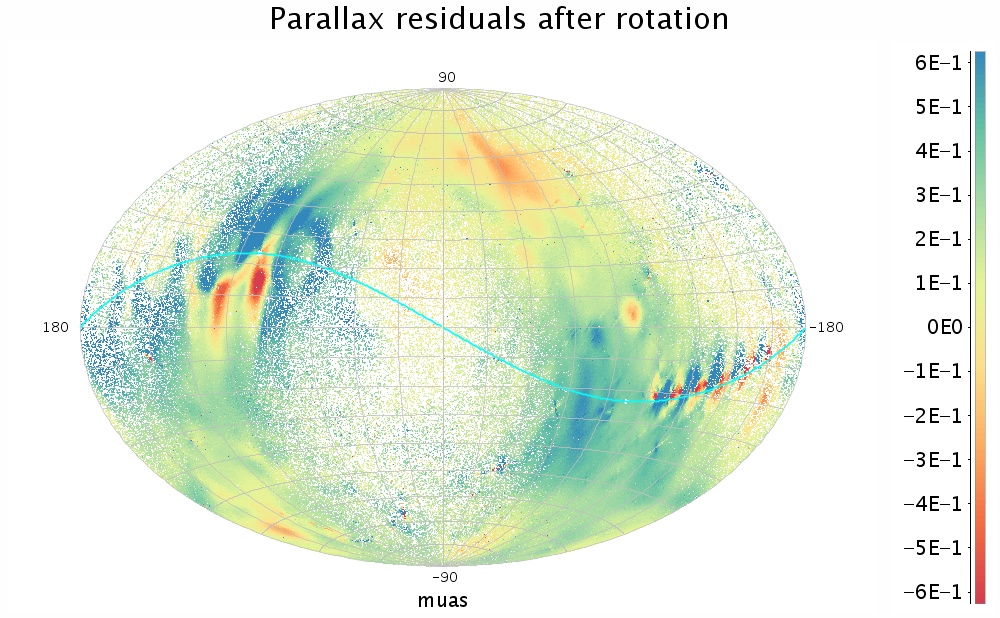}\hfill{}}
	\resizebox{\hsize}{!}
    {\includegraphics{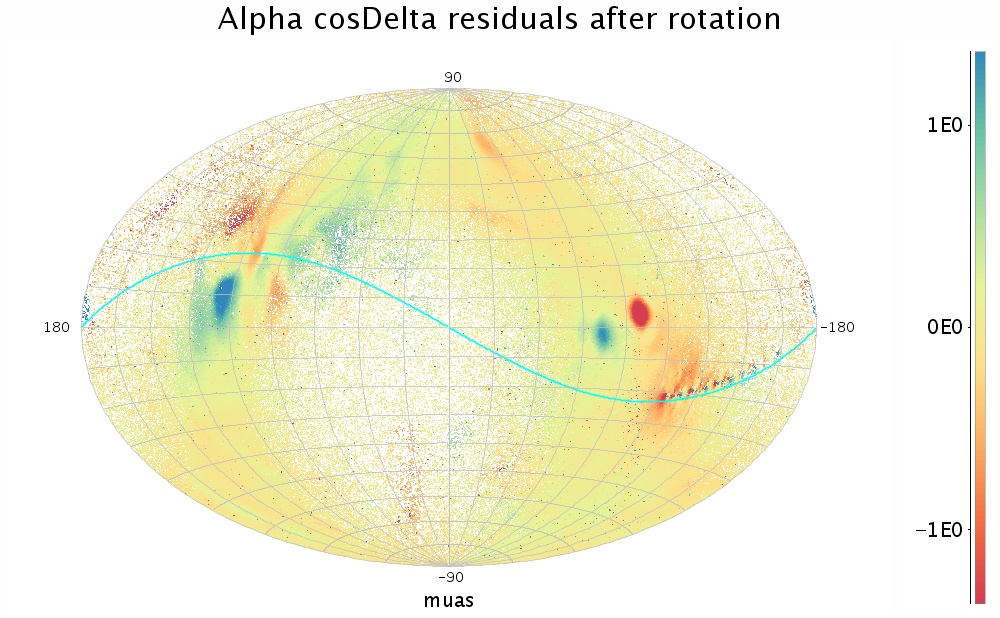}\hfill{}\includegraphics{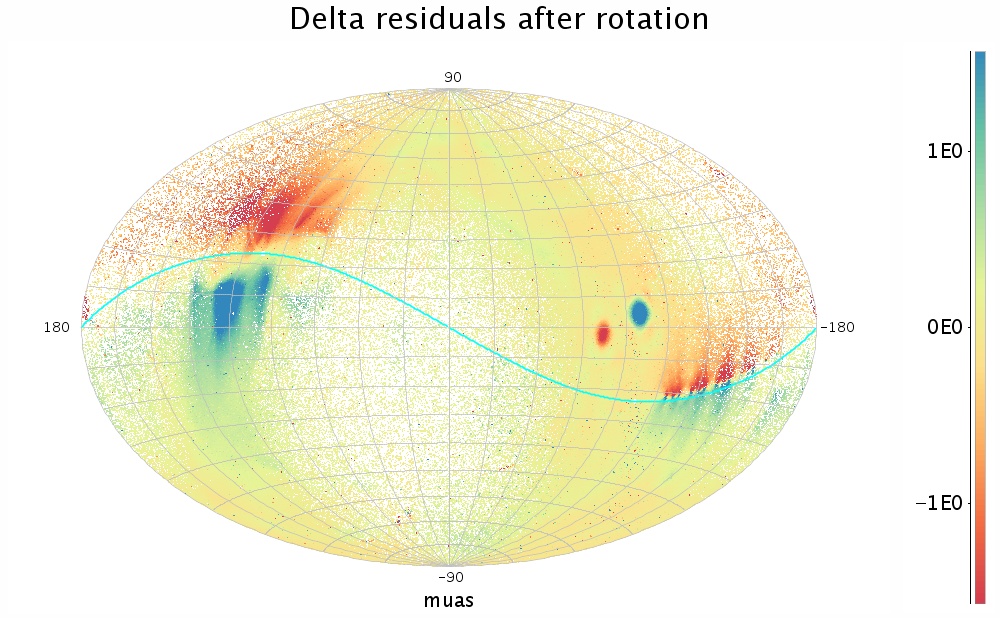}}
	\resizebox{\hsize}{!}
    {\includegraphics{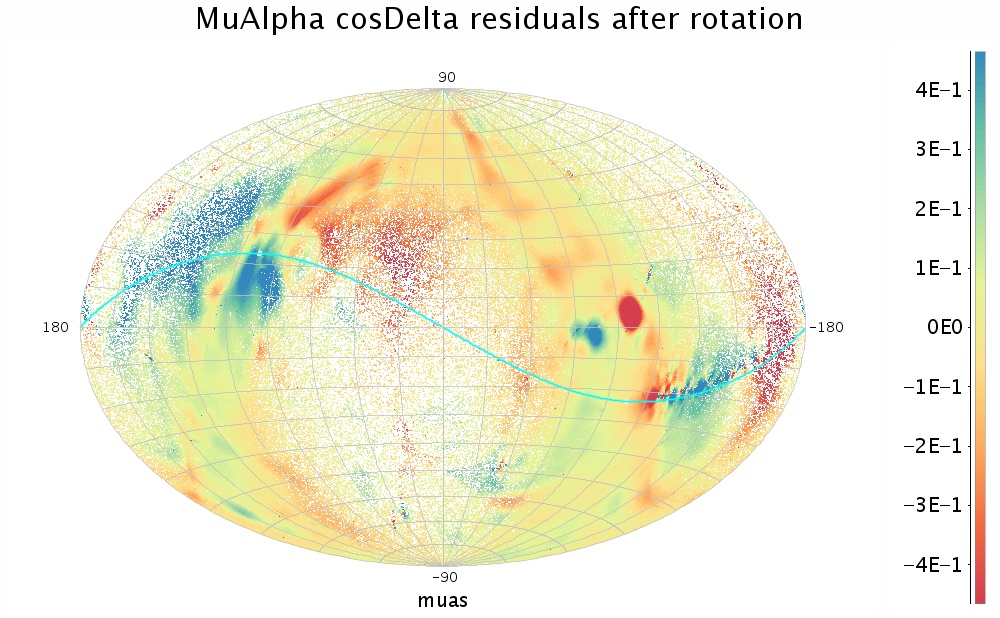}\hfill{}\includegraphics{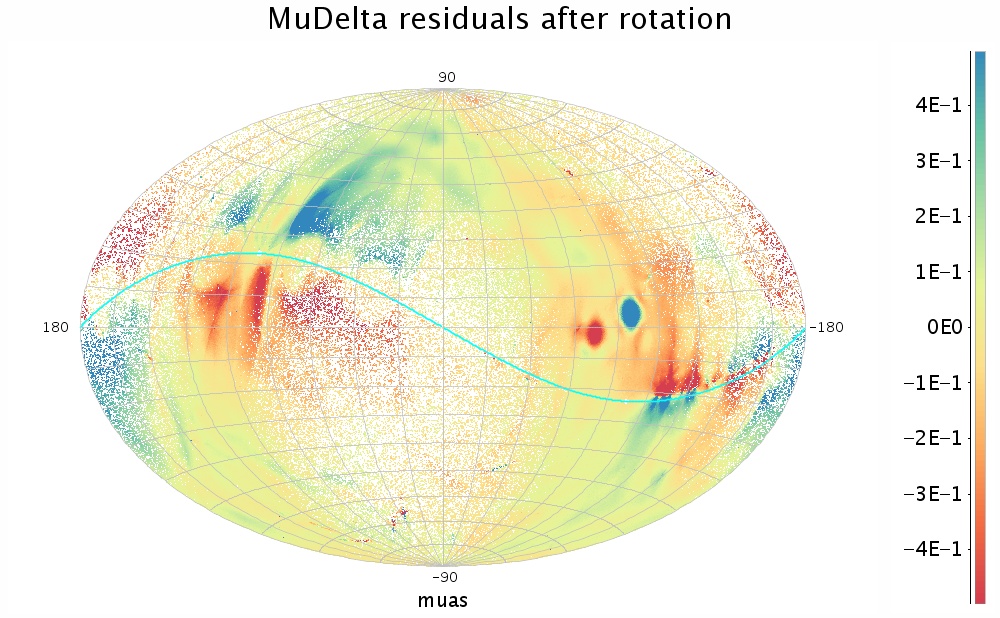}}
     \caption{All-sky map of the astrometric residual differences (in $\mu$as or $\mu$as$/$yr) for the noise-free GSR solution after the third step.}
     \label{fig:Astrometric_plots-1}
\end{figure*}

\paragraph{Astrometric parameters.}
The first-step solution of the test reflects the effects of the mismodelling on the accuracy level attainable with the current RAMOD2d model. The solution shows how this model, at first, reaches a systematic floor of $\sim10~\mu$as that, as expected given the noise-free observations, is independent from the magnitude range of the stars.

The second step provides a solution similar to the previous one because it starts from the same input catalog. Nonetheless, in this intermediate passage the BA modulation is removed from the measurements using its reconstruction from the first step. Although statistically similar to the previous one, in this way one obtains a better starting point for the final step. Indeed, jumping directly from the first to the third step, namely using the solution obtained without this ``preemptive cleaning'' of the known terms, would result in a less accurate final solution.

The third step, finally, gives the confirmation that the errors previously obtained are really the effect of modelling errors, since in this case the results are at the sub-$\mu$as level or better, as expected. This is also a confirmation that the adopted special procedure is able to recover the AGIS numerical accuracy, which implies that the same expectations have to apply to the second test.

Table~\ref{tab:Astrometric-results-gsr_test-1} reports the results of the first and third steps. The second step is not reported because of the negligible statistical differences with respect to the first one. Following the methodology of \citet{2012A&A...538A..78L} we adopted the Robust Scatter Estimation (RSE\footnote{The RSE is defined as 0.390152 times the difference between the 90th and 10th percentiles of the dataset. For a Gaussian distribution it coincides with the standard deviation.}) as a measure of the errors of the solution with respect to the true simulated values. It is worth noting that the medians of the parallaxes of the third step are one order of magnitude larger than those of the other unknowns. This residual noise can be neglected, as it is safely below the Gaia level of accuracy. Since prelilminary results have shown that with the implementation of a full-accuracy astrometric model this feature disappears \citep{2017A&A...608A..83B}, it can be attributed to a residual effect of the accuracy of the GSR astrometric model.

The all-sky plots of the astrometric solution for the third step are shown in Figure~\ref{fig:Astrometric_plots-1}. It is worth discussing in more detail the residual differences of these parameters. In general the largest differences show up close to the ecliptic, which is consistent with the fact that the lack of accuracy of RAMOD2d is basically due to imperfect modeling of the planets' contribution to the null geodesic. Such discrepancy can be seen, in a plot of the GREM vs. RAMOD differences between the measurements, as a sign flip when the observing direction crosses the position of a planet. This sign flip translates in a corresponding sign flip of the astrometric parameters. Since the planets move along the ecliptic, this systematic effect is smoothed out along this plane, and we only see the sign flip in the orthogonal direction.

\begin{table}
\caption{Attitude results (in $\mu\mathrm{as}$) for test 1.}
\centering
\label{tab:Attitude-results-1}
	\begin{tabular}{c c c c c }
	\hline\hline
	 &  & $\mathbf{e}_{1}$ & $\mathbf{e}_{2}$ & $\mathbf{e}_{3}$ \tabularnewline
	\hline 
	\multirow{2}{*}{1st step} & Average & 0.06 & -1.19 & 0.41 \tabularnewline
	 & RSE & 8.88 & 12.63 & 7.93 \tabularnewline
	\hline 
	\multirow{2}{*}{3rd step} & Average & 0.00 & -1.14 & 0.02 \tabularnewline
	 & RSE & 1.11 & 1.05 & 0.25 \tabularnewline
	\hline 
	\end{tabular}
\end{table}

\paragraph{Attitude Parameters.}
Similar considerations hold for the attitude parameters, as shown in Table~\ref{tab:Attitude-results-1}, where the numbers give the residual rotations around the three axes in $\mu$as. It is interesting to notice a residual, $\mu$as-level average for the $\mathbf{e}_{2}$ axis which, as in the case of the parallaxes, is linked to the accuracy of the astrometric model.

\begin{figure}
     \includegraphics[width=\hsize]{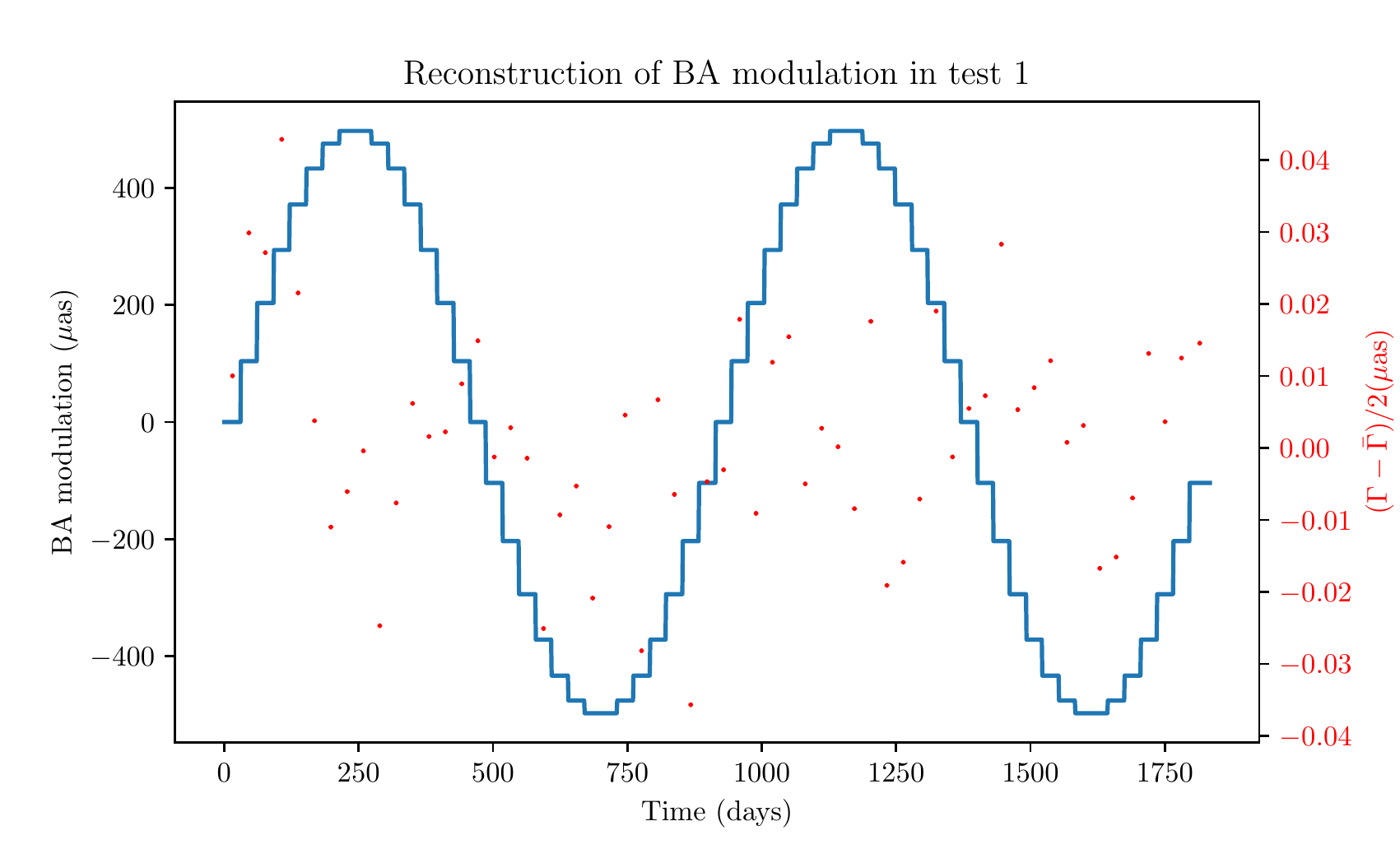}
     \caption{Test 1 BA reconstruction for FoV1. That of FoV2 is not reported as it is the same with the opposite sign. The blue line represents the true modulation signal, while the red dots, which use the scale on the right side of the plot, are the differences between such signal and the final reconstruction after the three steps.}
     \label{fig:BAV-1}
\end{figure}

\paragraph{Reconstruction of the Basic Angle Modulation.}
Figure~\ref{fig:BAV-1} shows the results of the BA modulation reconstruction for FoV1 as obtained from the $L_0$ (shift) $\Delta\eta$ large scale instrument parameters. Those of FoV2 are exactly antisymmetric. The residuals are at the sub-sub-$\mu$as level (with average $-0.0016\,\mu$as and stdev $0.015\,\mu$as for FoV1 and $0.0014\,\mu$as and stdev $0.015\,\mu$as for FoV2).

\subsection{Test 2: realistic solution with noisy observations}
\begin{figure*}
	\resizebox{\hsize}{!}
	{\includegraphics{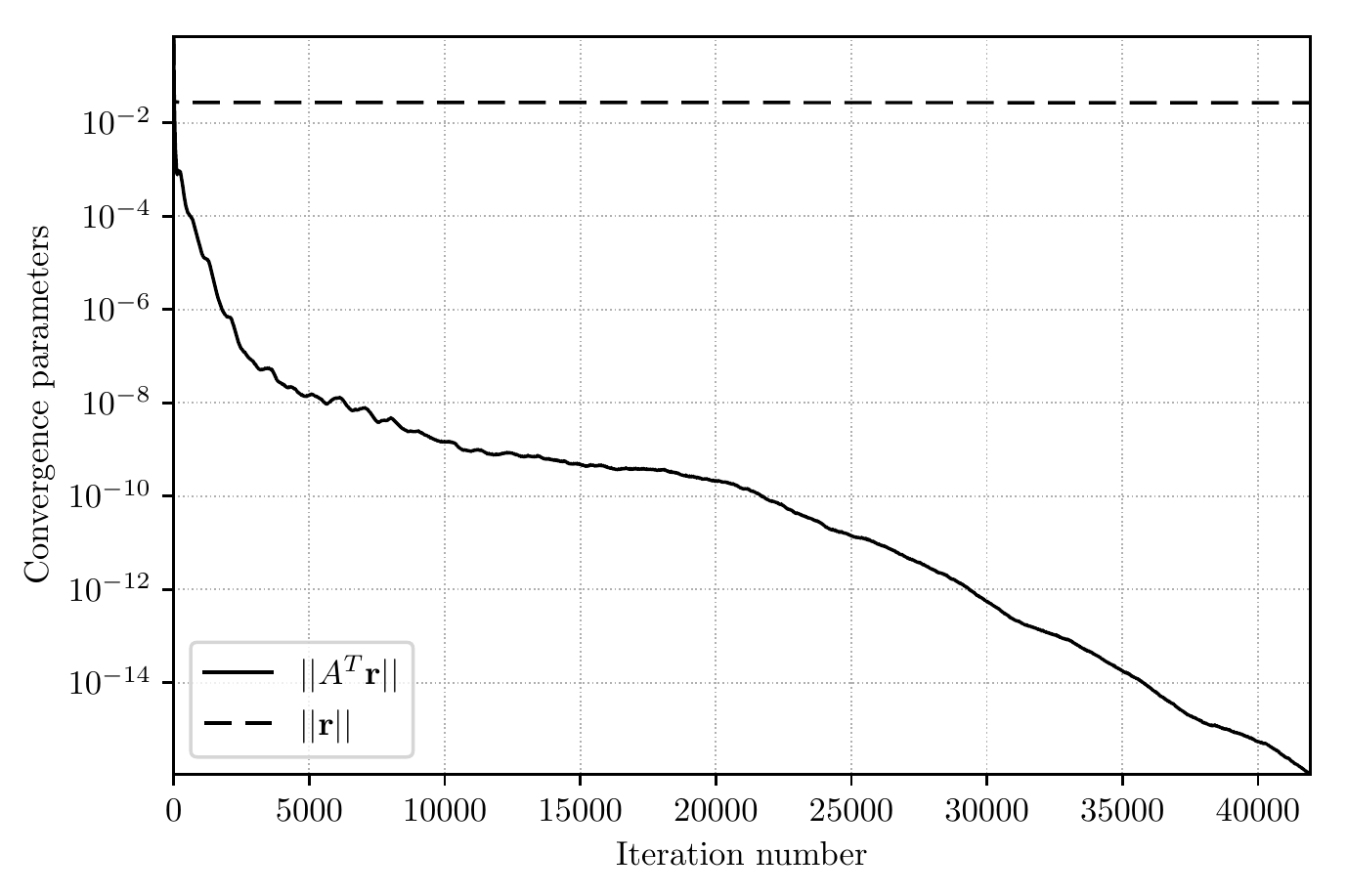}\hfill{}\includegraphics{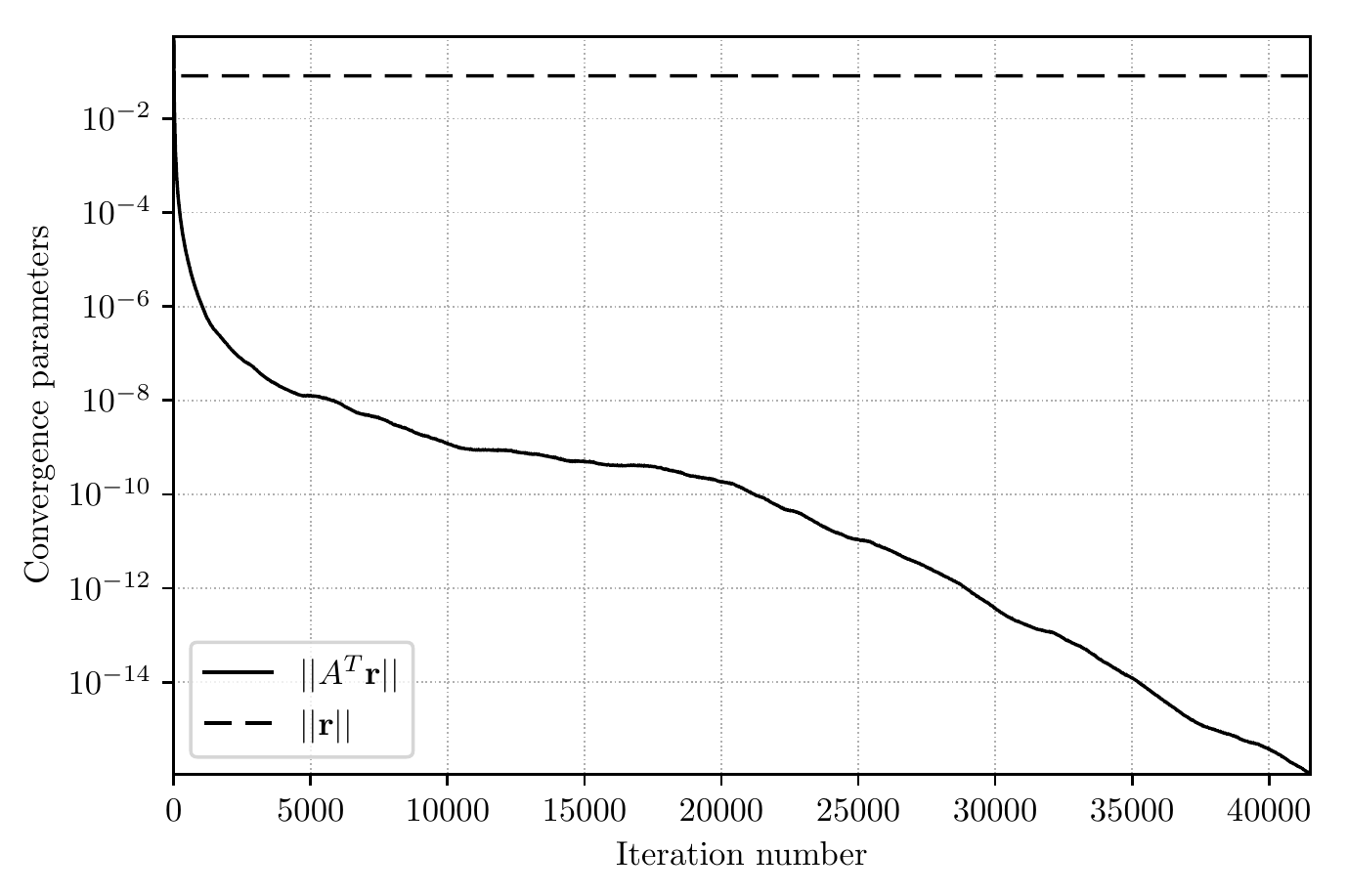}}
	\caption{\label{fig:Convergence-plots-2}Convergence plots of the first and third step of test 2.}
\end{figure*}

As for the previous test, the GSR pipeline filters out 460 stars that cannot be solved because of the requirements explained in the previous section. This leaves a system with $908\,519$ stars and just a slightly different number ($660\,594$) of degrees of freedom for the attitude. The large scale instrument parameters that are estimated in the solution are associated to the same total number of 63 CCDs, each varying with a time scale of one month. The number of astrometric unknowns is therefore $4\,542\,595$, that of the attitude parameters is $1\,981\,782$, and nothing changes for the instrument parameters (they are $45\,360$ as before). The system is thus described by a design matrix of $6\,569\,737$ columns (unknowns) and $1\,330\,628\,449$ rows (observations). In each of the three steps of the demonstration run the solver converged to the machine-precision least-squares solution in something less than $42\,000$ iterations, with a condition number of $\sim10^{6}$ (Figure~\ref{fig:Convergence-plots-2}).

\begin{table*}
\caption{Astrometric results (estimated minus true) for test 2. Units are in $\mu\mathrm{as}$ for parallaxes and positions and $\mu\mathrm{as}/\mathrm{yr}$ for proper motions. Right ascension and the corresponding proper motion are provided as $\alpha^*=\alpha\cos\delta$ and $\mu_{\alpha^*}=\mu_\alpha\cos\delta$. The RSEs of the AGIS demonstration run as reported in \citet{2012A&A...538A..78L} are shown for an easier comparison.}
\label{tab:Astrometric-results-gsr_test-2-vs-AGIS}
\centering          
\begin{tabular}{c c c c c c c c c c c c }     
\hline\hline       
  \multirow{2}{*}{Magnitude range} & \multirow{2}{*}{Step} & \multicolumn{5}{c}{Median} & \multicolumn{5}{c}{RSE} \\
  & & $\varpi$ & $\alpha^{*}$ & $\delta$ & $\mu_{\alpha^{*}}$ & $\mu_{\delta}$ & $\varpi$ & $\alpha^{*}$ & $\delta$ & $\mu_{\alpha^{*}}$ & $\mu_{\delta}$ \\
\hline
  \multirow{3}{*}{$\phantom{13\leq}G<13$} & 1 & -0.9 & 3.1 & 1.8 & 3.7 & -0.8 & \phantom{00}8.5 & \phantom{0}13.1 & \phantom{00}9.8 & \phantom{0}14.2 & \phantom{00}7.6 \\
  	& 3 & 0.2 & -0.3 & 0.1 & -0.4 & 0.1 & \phantom{00}7.9 & 7.0 & \phantom{00}6.2 & 4.9 & \phantom{00}4.4 \\
    & AGIS & - & - & - & - & - & \phantom{00}7.5 & 6.6 & \phantom{00}5.7 & 4.5 & \phantom{00}4.0 \\
  \cline{2-12}
  \multirow{3}{*}{$13\leq G<15$} & 1 & -0.9 & 2.3 & 2.4 & 3.1 & -0.8 & \phantom{0}14.8 & \phantom{0}17.0 & \phantom{0}13.4 & \phantom{0}16.4 & \phantom{0}10.4 \\
  	& 3 & 0.0 & -0.3 & 0.1 & -0.3 & 0.0 & \phantom{0}14.4 & \phantom{0}11.8 & \phantom{0}10.5 & \phantom{00}8.4 & \phantom{00}7.4 \\
    & AGIS & - & - & - & - & - & \phantom{0}14.9 & \phantom{0}12.4 & \phantom{0}10.6 & \phantom{00}8.7 & \phantom{00}7.5 \\
  \cline{2-12}
  \multirow{3}{*}{$15\leq G<16$} & 1 & -0.9 & 1.3 & 3.0 & 2.3 & -0.7 & \phantom{0}25.3 & \phantom{0}24.6 & \phantom{0}20.2 & \phantom{0}21.7 & \phantom{0}15.0 \\
  	& 3 & 0.1 & -0.5 & 0.1 & -0.4 & 0.0 & \phantom{0}25.1 & \phantom{0}20.2 & \phantom{0}17.9 & \phantom{0}14.2 & \phantom{0}12.6 \\
    & AGIS & - & - & - & - & - & \phantom{0}24.9 & \phantom{0}20.2 & \phantom{0}17.3 & \phantom{0}14.3 & \phantom{0}12.3 \\
  \cline{2-12}
  \multirow{3}{*}{$16\leq G<17$} & 1 & -1.0 & 0.7 & 3.1 & 1.8 & -0.4 & \phantom{0}40.2 & \phantom{0}35.2 & \phantom{0}29.7 & \phantom{0}29.0 & \phantom{0}21.8 \\
 	& 3 & -0.2 & -0.6 & 0.0 & -0.3 & 0.1 & \phantom{0}40.0 & \phantom{0}31.9 & \phantom{0}27.7 & \phantom{0}22.5 & \phantom{0}19.7 \\
    & AGIS & - & - & - & - & - & \phantom{0}38.4 & \phantom{0}30.8 & \phantom{0}26.7 & \phantom{0}21.8 & \phantom{0}19.0 \\
  \cline{2-12}
  \multirow{3}{*}{$17\leq G<18$} & 1 & -1.0 & 0.1 & 3.6 & 1.2 & -0.1 & \phantom{0}65.5 & \phantom{0}54.6 & \phantom{0}46.0 & \phantom{0}41.9 & \phantom{0}33.5 \\
 	& 3 & -0.2 & -0.4 & 0.1 & -0.1 & -0.1 & \phantom{0}65.4 & \phantom{0}52.4 & \phantom{0}44.9 & \phantom{0}36.7 & \phantom{0}32.0 \\
    & AGIS & - & - & - & - & - & \phantom{0}61.8 & \phantom{0}49.4 & \phantom{0}42.8 & \phantom{0}34.8 & \phantom{0}30.4 \\
  \cline{2-12}
  \multirow{3}{*}{$18\leq G<19$} & 1 & -0.9 & 0.1 & 3.9 & 0.5 & -0.2 & 110.8 & \phantom{0}90.3 & \phantom{0}76.1 & \phantom{0}65.7 & \phantom{0}54.8 \\
 	& 3 & -0.1 & 0.1 & 0.6 & -0.1 & -0.2 & 110.8 & \phantom{0}89.0 & \phantom{0}75.4 & \phantom{0}62.4 & \phantom{0}53.8 \\
    & AGIS & - & - & - & - & - & 104.1 & \phantom{0}83.3 & \phantom{0}70.7 & \phantom{0}58.9 & \phantom{0}50.8 \\
  \cline{2-12}
  \multirow{3}{*}{$19\leq G\phantom{<15}$} & 1 & -0.2 & -0.3 & 2.8 & 0.1 & -0.4 & 199.7 & 161.5 & 134.3 & 115.3 & \phantom{0}95.8 \\
 	& 3 & 0.3 & 0.4 & -0.3 & 0.1 & -0.5 & 199.5 & 160.8 & 134.0 & 113.6 & \phantom{0}95.4 \\
    & AGIS & - & - & - & - & - & 207.6 & 167.9 & 140.0 & 118.5 & 100.2 \\
\hline
\end{tabular}
\end{table*}

\paragraph{Astrometric parameters.}
Table~\ref{tab:Astrometric-results-gsr_test-2-vs-AGIS} reports, for each magnitude class, the median and the RSE for steps 1 and 3 of the GSR runs, as well as the corresponding results of the AGIS demonstration run.

As anticipated above, the first step produces an improved astrometric catalog, which however is still affected by systematics; $\mu$as-level medians and ratios of the $\varpi$ errors to those of the other astrometric unknowns different from expectations  are the numerical signatures of such systematics. In particular, it is known from both theoretical analysis and numerical simulations that the scanning law of Gaia should produce well-defined values for these ratios, specifically,
\begin{equation}
\bar{r}_{\alpha^{*}}\equiv\frac{\sigma_{\alpha^{*}}}{\sigma_{\varpi}}=0.787,
\end{equation}
and similarly $\bar{r}_{\delta}=0.699$, $\bar{r}_{\mu_{\alpha^{*}}}=0.556$, $\bar{r}_{\mu_{\delta}}=0.496$ (\url{https://www.cosmos.esa.int/web/gaia/science-performance}). These approximate estimations can thus be used as an additional check for the solutions, and the results from Table~\ref{tab:Astrometric-results-gsr_test-2-vs-AGIS} show that the actual ratios are close to the expected values, and similar between them, for AGIS and for the step 3 of the GSR solution, whereas this is not the case for the step 1.

\begin{figure*}
	\resizebox{\hsize}{!}
	{\includegraphics{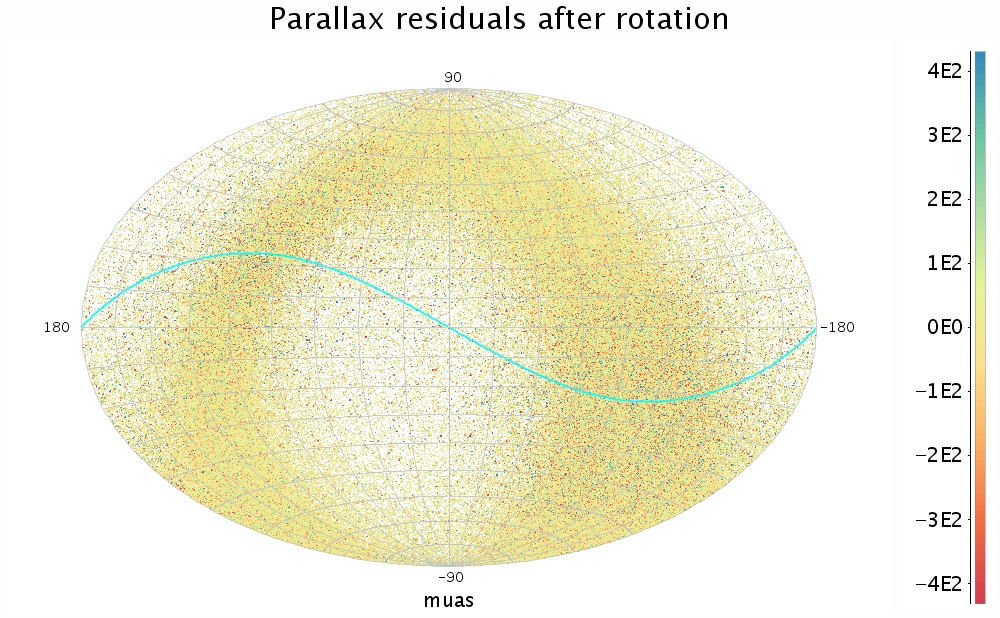}\hfill{}\includegraphics{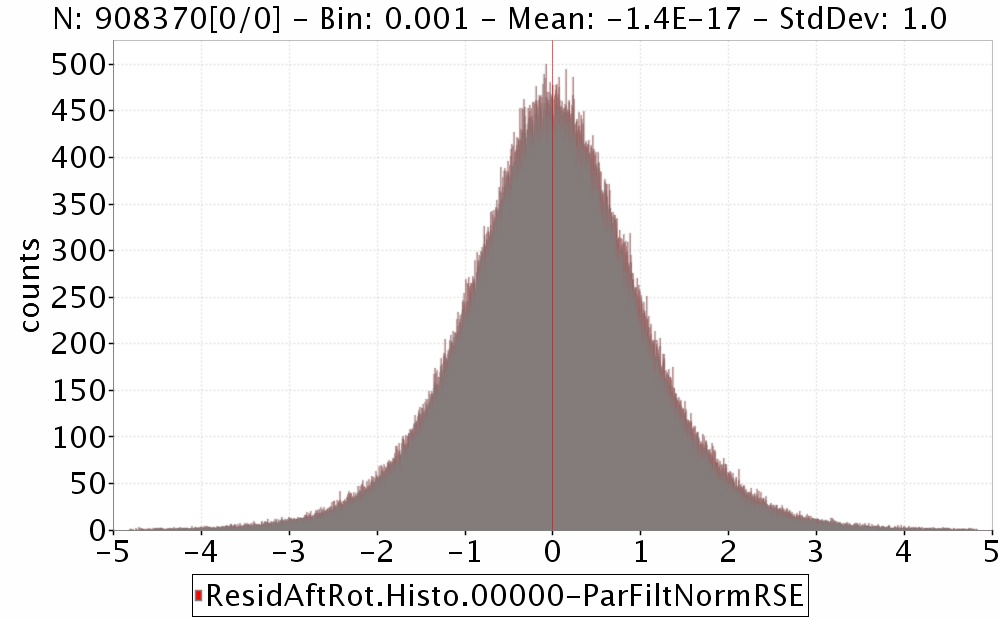}}
	\caption{\label{fig:Astrometric_plots-2}All-sky map and normalized histogram of the parallax astrometric residual differences for the GSR solution after the third step. Units of the all-sky map are $\mu$as.}
\end{figure*}

For test 2, the all-sky map plots do not show the systematics of the previous test. Indeed, in this test the remaining sub-$\mu$as systematic errors visible in the previous plots are completely negligible with respect to the Gaussian ones, which distribute uniformly on the sky. The plots for all the unknowns are quite similar, so in Figure~\ref{fig:Astrometric_plots-2} we just show the case of the parallaxes, where the all-sky map is paired with the histogram of the normalized astrometric errors obtained by combining the data from each magnitude class. These data show that the solution of the third step, eventually, recovers the remaining modelling residuals, and that with this ad-hoc procedure GSR is able to recover an astrometric solution comparable to that of AGIS at the sub-$\mu$as level even with a less accurate astrometric model.

\begin{table}
\caption{Attitude results (in $\mu\mathrm{as}$) for test 2.}
\centering
\label{tab:Attitude-results-2}
	\begin{tabular}{c c c c c c }
	\hline\hline
      &  & $\mathbf{e}_{1}$ & $\mathbf{e}_{2}$ & $\mathbf{e}_{3}$ & $\mathbf{e}_{2}/\mathbf{e}_{1}$ \\
    \hline
    \multirow{2}{*}{1st step} & Average & -0.12 & -0.91 & 0.19 & - \\
      & RSE & 307.76 & 411.52 & 25.76 & 1.34 \\
    \hline 
    \multirow{2}{*}{3rd step} & Average & -0.19 & -1.80 & -0.12 & - \\
      & RSE & 308.55 & 411.12 & 24.02 & 1.33 \\
    \hline 
    \multirow{2}{*}{AGIS} & Average & - & - & - & - \\
      & RSE & 167 & 224 & $\sim20$ & 1.34 \\
	\hline 
	\end{tabular}
\end{table}

\paragraph{Attitude parameters.}
\begin{figure}
	\includegraphics[width=\hsize]{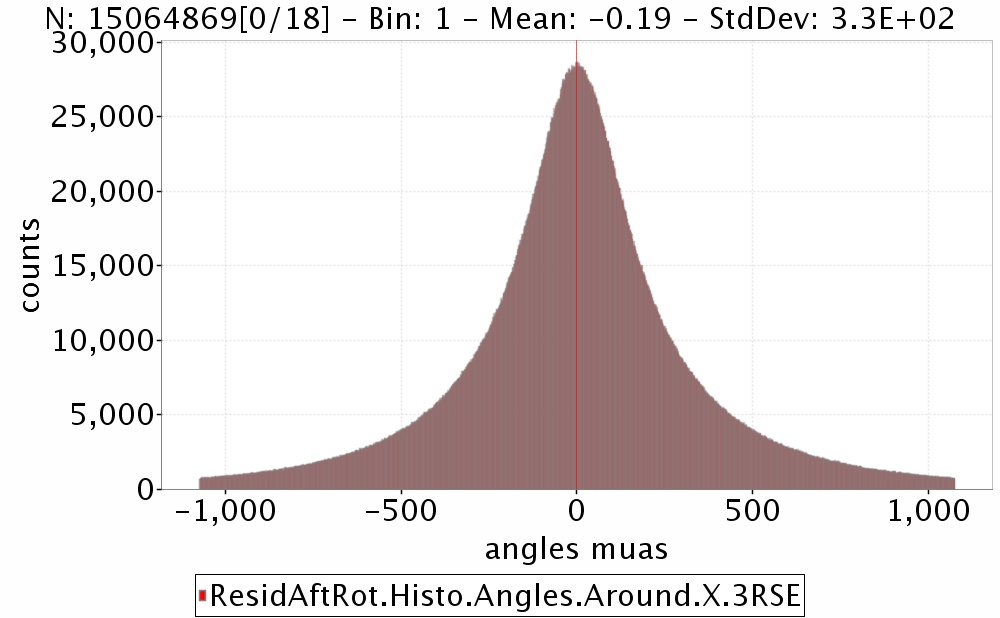}
    \includegraphics[width=\hsize]{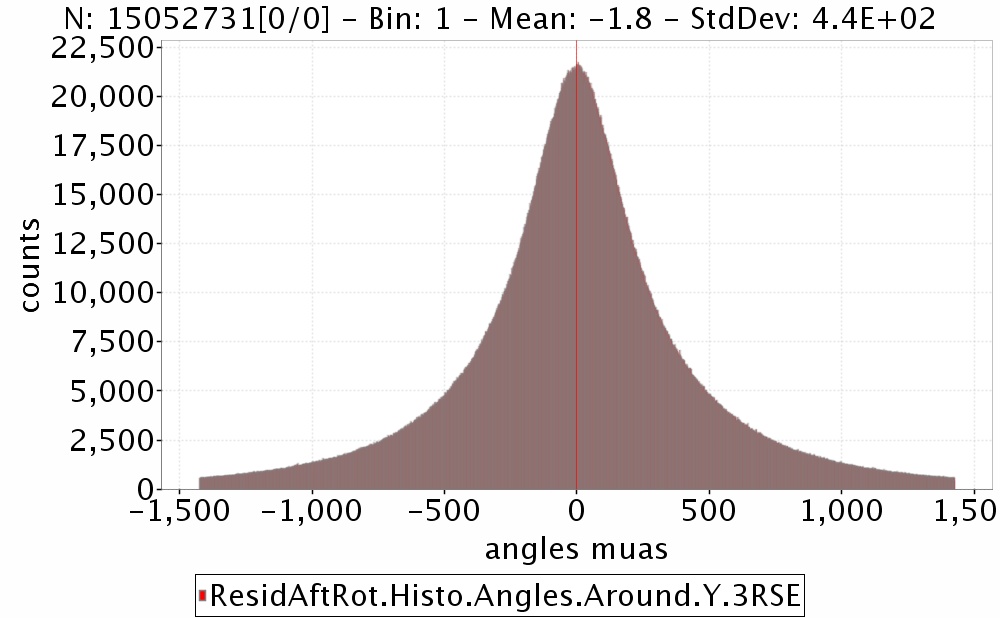}
	\includegraphics[width=\hsize]{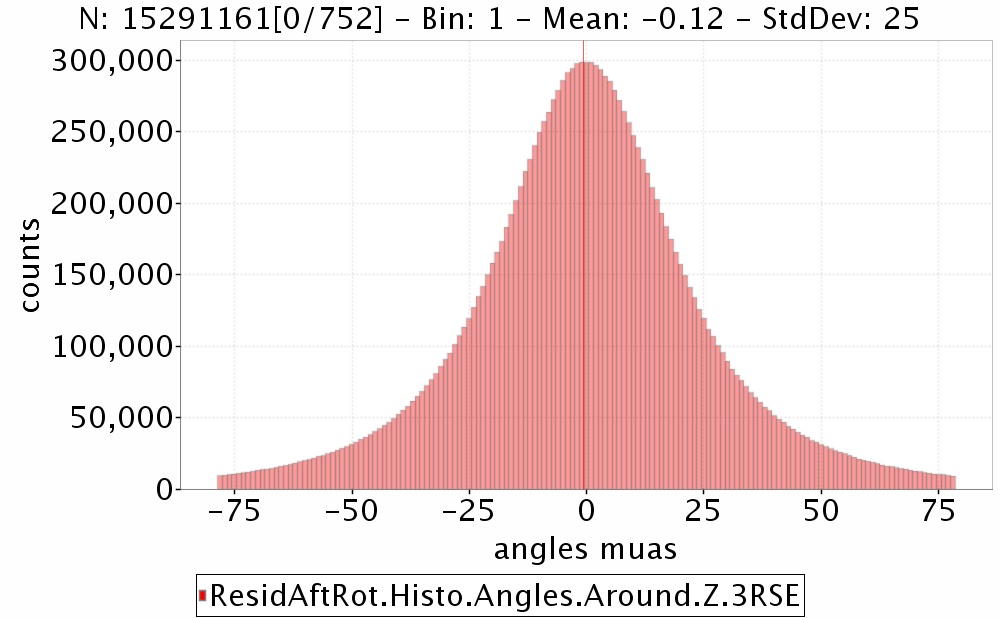}
	\caption{\label{fig:attitude_histo-2}Histograms of the attitude solution after the third step of test 2 for the $x$, $y$ and $z$ axes.}
\end{figure}

Similarly, we report in Table~\ref{tab:Attitude-results-2} the results for the attitude solutions from the first and third step of GSR, and those of AGIS. The numbers give the residual rotations around each axis in $\mu$as. The last column reports the ratio between the residual errors of $\mathbf{e}_{2}$ and $\mathbf{e}_{1}$ ($y$ and $x$ axes respectively), which from the geometry of the observations should be 1.34.

While the final accuracy of the astrometric parameters mostly depends on the number of observations per star, that of the attitude strongly depends on the total number of observations, and therefore on the number of stars involved in the sphere reconstruction and their magnitude. For this reason GSR, with $908\,000$ stars, cannot reach the same accuracy obtained by AGIS with about $2\,300\,000$ stars. The degradation factor must be close to $1.6$, which compares well to the $1.8$ obtained in this test. Finally, Figure~\ref{fig:attitude_histo-2} shows the histograms of the attitude errors for the $x$, $y$ and $z$ axes.

\begin{figure}
     \includegraphics[width=\hsize]{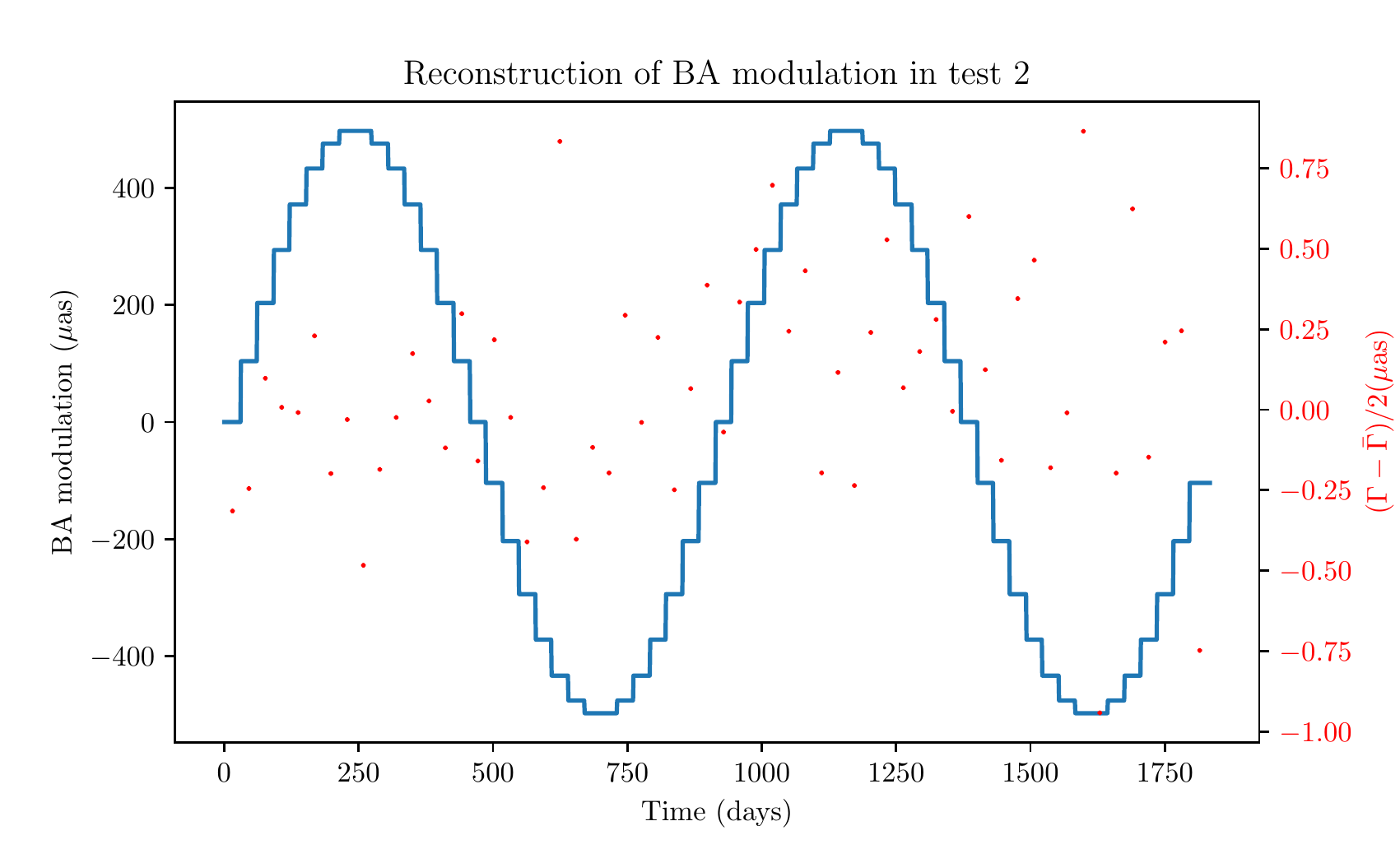}
     \caption{Test 2 BA reconstruction for FoV1. That of FoV2 is not reported as it is the same with the opposite sign. As for the previous test, the blue line represents the true modulation signal, while the red dots, which use the scale on the right side of the plot, are the differences between such signal and the final reconstruction after the three steps.}
     \label{fig:BAV-2}
\end{figure}

\paragraph{Reconstruction of the basic angle modulation.}
Figure~\ref{fig:BAV-2} shows the results of the BA modulation reconstruction for FoV1 as obtained from the $L_0$ $\Delta\eta$ large scale instrument parameters. Those of FoV2 are exactly antisymmetric. As in the case of AGIS, the errors are at the sub-$\mu$as level (with average $-0.053\,\mu$as and stdev $0.357\,\mu$as for FoV1 and $-0.047\,\mu$as and stdev $0.361\,\mu$as for FoV2).

\begin{table}
\caption{AL and AC standard deviations of the residuals of the non-weighted equation system (``System'' columns) after step 3 of test 2 per magnitude class compared with the single-measurement error used to generate the simulated data (``Est.'' columns).}
\centering
\label{tab:obsres_step3}
	\begin{tabular}{c r r r r }
      \hline\hline
      \multirow{2}{*}{Magnitude range} & \multicolumn{2}{c}{AL $\sigma$} & \multicolumn{2}{c}{AC $\sigma$} \\
       & Est. & System & Est. & System \\
      \hline
      $\phantom{13\leq}G<13$  &   76 &   78 &   348 &   400 \\
      $13\leq G<15$           &  175 &  180 &   809 &   930 \\
      $15\leq G<16$           &  310 &  320 &  1472 &  1700 \\
      $16\leq G<17$           &  495 &  510 &  2485 &  2900 \\
      $17\leq G<18$           &  801 &  840 &  4494 &  5200 \\
      $18\leq G<19$           & 1133 & 1400 &  8969 & 10000 \\
      $19\leq G\phantom{<19}$ & 2345 & 2700 & 19695 & 23000 \\
      \hline 
	\end{tabular}
\end{table}

\paragraph{Observations' residuals.} The solution obtained after the third step is used to compute the AL and AC residuals of the condition equations. As long as the solution is close to the true values, the standard deviation of the distribution of the residuals for each star should approach the Gaussian noise used to simulate the measurement error. Table~\ref{tab:obsres_step3} reports such residuals for each magnitude class, along with the single-measurement error approximate values for the simulated data estimated with the ad-hoc empirical formulae used for Figure~\ref{fig:GaiaAccEst}. It has to be stressed that the values obtained with the empirical formulae are computed under the hypothesis of a Gaussian distribution with zero average for any object. For the actual residuals, the least-squares solution implies a close-to-zero average for the complete system only, and nothing can be said in general for subsets of observation equations. This means that the AL and AC residuals of each star will be generally distributed around a non-zero average, therefore the estimation by the empirical formula binned by magnitude class is likely to underestimate the real case, which is exactly what we observe in the table.

On the other hand, as we have anticipated above, we have to expect close-to-zero residuals for the complete system we actually solved (in the least-squares sense) that is the one weighted by the measurement errors via the weight matrix $W$, $WA\mathbf{x}=W\mathbf{b}$. For this system we have to expect that $\langle W(A\mathbf{x}-\mathbf{b})\rangle\simeq0$. The actual result in this case is $-0.50~\mu\mathrm{as}$, which is again at the sub-$\mu\mathrm{as}$ level as expected from the accuracy of the solution as compared to the true values.

\section{Current status and future developments}
Work is currently ongoing to further develop the AVU/GSR pipeline. In the following we detail the main issues that are going to be addressed in future developments.

\subsection{Convergence speed and variance estimation.} As shown in the previous section, the number of iterations needed to reach a complete convergence can be very different for different situations. It is well known that such variation depends on the number of unknowns of the problem, however experience is showing, as it is clear already from the results of Test 1 and Test 2, that other factors can have a strong influence on the convergence speed. For example, this demonstration run used the Calibration parameters only to estimate a very specific signal injected in the BA, but it is expected that, when the full instrument model will be adopted, an even larger number of iterations will be needed to reach the convergence. It would be obviously an advantage to minimize such number.

One possibility is to use different kind of astrometric constraints. The full astrometric problem, unlike that represented by the AGIS approach, is intrinsically ill-conditioned, and its solution requires the introduction of six constraint equations that fix the orientation of the reference system for the positions and the proper motions. As explained in Sect.~\ref{sec:DemonRun}, currently GSR implements such equations by constraining the values of one-and-a-half stars, but other criteria are applicable. For example, it is possible to compute six ``barycentric constraints'' using the values of an arbitrary selection of stars. The determination of such constraints with a large set of stars usually reduces the iterations needed for the convergence.

Another possibility is simply to relax the convergence conditions. In order to guarantee the best possible solution from a purely numerical point of view, it was decided to set the convergence requirement to the most stringent value, that is to machine-precision level. However, it has been noticed that the solution $\delta\mathbf{x}$ stabilizes before the convergence parameters reach such level, and it was verified that setting it, for example, to the less demanding value of $\sim10^{-13}$ would have produced an equivalent $\delta\mathbf{x}$ with a significantly reduced number of iterations. On the other hand, there is no obvious way to link the stability of the solution at the required numerical accuracy with the  numerical value of $||A^\mathrm{T}\mathbf{r}||$, namely the residuals of the normal system that is used to estimate the convergence status of non-compatible systems. A possible practical solution is to check at regular intervals whether the variation of the solution falls below a given threshold. Technically, this option is easy to implement, since it can take advantage of a feature of our customized implementation of the LSQR algorithm, which produces intermediate solutions at regular intervals of iterations. Another option for the threshold value of the convergence condition is $||A-\bar{A}||/||A||$, as suggested in \citet{PaigeSaunders1982}, in which $\bar{A}$ are the true values of the coefficients' matrix and the numerator can be estimated by knowing the uncertainties of the catalog values. Both these options, however, require further study to assess the reliability of the results obtained in this way. A drastic improvement in this respect, however, might have an undesired consequence.

Actually, it is also known that initially the LSQR algorithm underestimates the variances of the solution, and that such estimation improves with the number of iterations. A careful trade-off is therefore needed to get the fastest possible convergence without hampering the variances estimation or, in alternative, one might decide to resort to other approximate methods to estimate the latter. The GSR implementation of the LSQR algorithm always provides the variances along with the solution, but currently the former are used just for a first rough estimation of the quality of the solution. Further work is needed to assess the reliability of the variances estimation.

\subsection{BA fit and calibration model.} As remembered in Sect.~\ref{sec:Modeling-the-observations}, the daily calibration does not guarantee a reconstruction of the instrument parameters at the final accuracy level required by the sphere reconstruction. Additional calibration parameters are thus added to the observation equations in order to recover any remaining un-calibrated residuals. However, the instrument model currently implemented in GSR is still inadequate to obtain a sphere solution coping with the accuracy goal of the Gaia catalog. Current work on real data is aimed at determining the improvements to the present instrument model needed to reach such an accuracy.

A similar reasoning has to be applied for the BA. The BAM pipeline(s) provides a daily reconstruction of the BA Variation (BAV), and it is likely that a cyclic reprocessing of these data can improve this reconstruction by providing a calibrated estimation of such variations. It is nonetheless appropriate to introduce further BA Correction (BAC) components in the observation equations to take into account a possible residual signal \citep{2018A&A...616A...2L}.

\subsection{Comparison analysis.} As described in Sect.~\ref{subsec:Comparison}, the GSR pipeline includes a Post-Solver module of comparison between the GSR and AGIS sphere reconstructions. The goal of this task is the internal validation of the primary star catalog astrometric parameters and associated formal uncertainties: this is carried out through a detailed analysis of the differences between the two solutions, thereby enlightening possible causes for discrepancies.

While the implementation of the VSH technique in the current GSR pipeline can be already used to identify and remove a residual rotation between the two catalogs, easily identified by the toroidal harmonics of degree one, the statistical robustness of the significance level test of other expansion coefficients, along with their interpretation in terms of plausible physical/geometrical effects, need a more extensive investigation, which is being addressed in the above cited forthcoming paper (\citealt{2019Comparison...InPrep..B}, in preparation).

Caution must be paid when the non-uniformity of the stellar distribution brakes the orthogonality of the base functions, with the consequence of giving rise to correlations among the VSH coefficients which must be taken into account; also, early truncation of the series expansion can generate spurious coefficients which result in false signal detection.

Another technique that we plan to adopt for the analysis of zonal errors is that of Infinitely Overlapping Circles (IOC, \citealt{1992ApJ...392..746T,1993JSCS...48...29B}): based on a statistical moving average naturally defined on the sphere; this technique is particularly easy to implement and can be successfully applied to extract local signals of scale length comparable or larger than the typical distance between neighboring sources.

\subsection{Full-accuracy relativistic models.} As shown in the present paper, GSR would benefit from the implementation of a relativistic model able to cope with the accuracy of the Gaia measurement. Work in this sense is in a very advanced stage, and will be reported in a forthcoming publication (\citealt{2019Models...InPrep..V}, in preparation).

\subsection{Handling outliers, observations' weighting and micro-events.} Observations' errors have to be used to build the weight matrix of the system. At present, this is done by using the expected uncertainty of the observed object (see Sect.~\ref{sec:LinEqSyst}) but, as explained in \citet{2012A&A...538A..78L}, a more effective weighting can be obtained by considering both these uncertainties and the actual observations' residuals, a procedure that in AGIS is also used to identify the outliers of the observations of a given source.

In addition to this, attitude undergoes uninterrupted perturbations because of the so-called micro-events, namely ``micro-clanks'' (small adjustments of the satellite structure) and ``micro-hits'' (due to the impact of micro-meteoroids with the Gaia satellite). These anomalies manifest themselves through non-nominal temporal variations of the AL and AC field angles, which can be interpreted as variations in the attitude scan rates. In the main pipeline they are estimated by a pre-processor running before the sphere solution, and successively removed from the observations. Failing to remove such perturbations would cause mas-level degradation to the accuracy of the sphere solution.

The above mentioned weighting procedure is currently under testing in GSR whereas, regarding the treatment of the micro-events, our pipeline is already able to remove them from the attitude corrections provided in the Gaia Main Database, while work is ongoing to implement GSR's own procedure for micro-clanks and micro-hits.

\section{Conclusions}
The Astrometric Verification Unit is in charge of providing the DPAC with a pipeline able to realize a reconstruction of the global astrometric sphere independent from that of AGIS. This would allow the double-checking of the determination of the global reference system of Gaia. To this aim it is sufficient to reproduce the first stage of the process implemented by AGIS, namely the solution of the global sphere from the so-called primary sources. The absence of the secondary sources in such a pipeline, named AVU/GSR, is compensated for by the comparison task, whose goal is to identify and characterize any systematic discrepancy between the AGIS and GSR solutions larger than the expected accuracy of the Gaia catalog.

In order to guarantee a sufficient independence between the two solutions, AVU/GSR uses a different relativistic model for the astrometric observable, and a different parametrization for the attitude, based on MRP rather than quaternions. Moreover, the solution of the system of linearized equations is performed with a parallelized implementation of the LSQR full-iterative algorithm.

In this paper we showed that GSR managed to successfully reproduce the results of the demonstration run of the AGIS pipeline, as illustrated in \citet{2012A&A...538A..78L}. This required the execution of two tests on simulated data for a 5-year mission duration, both solved for astrometric, attitude and instrument parameters.

The first one had the goal of assessing the sub-$\mu$as numerical accuracy of the GSR pipeline, which was obtained by using perturbed starting values and error-free observational data; the second one had to mirror the outputs of the AGIS demonstration run with perturbed values for source and attitude parameters, plus random measurement errors compatible with those expected by Gaia. Moreover, large-scale instrument calibrations were used to reconstruct a periodic signal injected in the BA value. This test was designed to produce an astrometric catalog with errors compatible with those of the final Gaia catalog.

Both tests performed according to expectations, even if the current implementation of the GSR relativistic model is less accurate than the AGIS one, its accuracy being limited by the influence solar system bodies other than the Sun on the computation of null geodesics. Despite these limitations, the current pipeline reaches the level needed by the Gaia measurements at appropriate elongation from the perturbing object. Moreover, the model is improved by finding approximate corrections to the light deflection effect induced by the planets and the Moon, thereby extending the required accuracy to larger portions of the sky.

The influence of this issue on the final solution was investigated in this work. Test 1 revealed that GSR is critically sensible to the catalog errors because of the limited accuracy of the model; at the same time it allowed to prove that the required quality of the solution can be reached by performing an external iteration. Test 2, instead, showed that, as it was expected, the manifestation of modelling errors depends on the accuracy of Gaia measurements, and therefore at the faint end of the stellar sample the AGIS and the GSR solutions are compatible already after the first iteration, while this is true for the entire catalog only after the external iteration. This special procedure will likely be unnecessary when the ongoing work of implementing an astrometric model at the same intrinsic accuracy of GREM will be completed.

This work has put to evidence that GSR is sufficiently mature to start processing real data, and current developments aim at providing this pipeline with the features needed to meet the accuracy requirements of the Gaia solution. Some of them are made necessary by the actual behavior of the instrument and have already been tackled by AGIS; others involve a more sophisticated treatment of the comparison task, and are specific to GSR. Aim of the GSR group is to have all the needed features implemented and tested for by the next Gaia Data Release, in order to team up with AGIS and thus contribute to the production of the Gaia catalog, as foreseen in the DPAC plans.

\begin{acknowledgements}
This work was supported by the Agenzia Spaziale Italiana (ASI) through contract 2014-025-R.1.2015 to the Italian Istituto Nazionale di Astrofisica (INAF) and contract 2016-17-I.0 to the Aerospace Logistics Technology Engineering Company (ALTEC S.p.A.), and INAF.

The authors wish to thank Michael Biermann (DPAC CU3 leader) and Gonzalo Gracia (DPAC Project Office Coordinator) for their support and advice during the evaluation of the results of the Demonstration Run, and the anonymous referees for their helpful comments.
\end{acknowledgements}

\bibliographystyle{aa}

\end{document}